\documentclass[letterpaper,10pt,journal,twoside]{IEEEtran}
   \usepackage[pdftex]{graphicx}

\usepackage[cmex10]{amsmath}
\usepackage{array}
\usepackage[numbers,sort&compress]{natbib}

\usepackage{multirow}
\usepackage[ruled,linesnumbered,vlined]{algorithm2e}
\usepackage{hyperref} 
\usepackage{booktabs}
\usepackage{colortbl}
\usepackage[flushleft]{threeparttable}
\usepackage{float}
\usepackage{array}
\usepackage{lettrine}
\usepackage{amsfonts}
\usepackage{bigstrut}
\usepackage{enumitem}
\setlist[description]{leftmargin=\parindent,labelindent=\parindent}

\usepackage{subfigure}
\usepackage[svgnames]{xcolor}

\usepackage{url}
\usepackage{hyperref}
\usepackage[hyphenbreaks]{breakurl}
\usepackage{colortbl}
\usepackage{caption}
\newcommand\VRule[1][\arrayrulewidth]{\vrule width #1}

\begin{document}

%
\title{Exploiting Spin-Orbit Torque Devices as Reconfigurable Logic for Circuit Obfuscation}
%
%
%
%

\author{Jianlei~Yang,~\IEEEmembership{Member,~IEEE,}
        Xueyan~Wang,~\IEEEmembership{Student~Member,~IEEE,}
        Qiang~Zhou,~\IEEEmembership{Member,~IEEE}
        Zhaohao~Wang,~\IEEEmembership{Member,~IEEE,}
        Hai~(Helen)~Li,~{Senior~Member,~IEEE} \\
        Yiran~Chen,~\IEEEmembership{Fellow,~IEEE}
        and~Weisheng~Zhao,~\IEEEmembership{Senior~Member,~IEEE}
\thanks{Manuscript received on May 2017, revised on July 2017, October 2017 and December 2017, accepted on January 2018. This work was supported in part by the National Natural Science Foundation of China (61602022, 61501013, 61571023, 61521091 and 1157040329), ZG216S1745, NSF CNS-1744111, Beijing Municipal of Science and Technology (No. D15110300320000), National Key Technology Program of China 2017ZX01032101 and the 111 Talent Program B16001. \textit{(Corresponding authors are Jianlei Yang and Weisheng Zhao.)}}
\thanks{J. Yang is with Fert Beijing Research Institute, BDBC, School of Computer Science and Engineering, Beihang University, Beijing, 100191, China. E-mail: jianlei@buaa.edu.cn} 
\thanks{ X. Wang and Q. Zhou are with Department of Computer Science and Technology, Tsinghua University, Beijing, 100084, China.} 
\thanks{ H. Li and Y. Chen are with Department of Electrical and Computer Engineering, Duke University, Durham, NC 27708, USA.} 
\thanks{ Z. Wang and W. Zhao are with Fert Beijing Research Institute, BDBC, School of Electronic and Information Engineering, Beihang University, Beijing, 100191, China. E-mail: weisheng.zhao@buaa.edu.cn}
}

%
%

\markboth{IEEE Transactions on Computer-Aided Design of Integrated Circuits and Systems}%
{Yang \MakeLowercase{\textit{et al.}}: Exploiting Spin-Orbit Torque Devices as Reconfigurable Logic for Circuit Obfuscation}
%



\maketitle

\begin{abstract}
Circuit obfuscation is a frequently used approach to conceal logic functionalities in order to prevent reverse engineering attacks on fabricated chips. Efficient obfuscation implementations are expected with lower design complexity and overhead but higher attack difficulties. In this paper, an emerging obfuscation approach is proposed by leveraging spin-orbit torque (SOT) devices based look-up-tables (LUTs) as reconfigurable logic to replace the carefully selected gates. It is essentially impossible to identify the obfuscated gate with SOTs inside according to the physical geometry characteristics because the configured functionalities are represented by magnetization states. Such an obfuscation approach makes the circuit security further improved with high exponential attack complexities. Experiments on MCNC and ISCAS 85/89 benchmark suits show that the proposed approach could reduce the area overheads due to obfuscation by 10\% averagely.
\end{abstract}

\begin{IEEEkeywords}
Spin-Orbit Torque, Magnetic Tunnel Junction, Reconfigurable Logic, Circuit Obfuscation.
\end{IEEEkeywords}

\IEEEpeerreviewmaketitle

\section{Introduction}\label{sec:introduction}

\subsection{Reverse Engineering of ICs/IPs}

\IEEEPARstart{W}{ith} the rapid increasing requirements of embedded systems and the internet of things (IoTs), application specific integrated circuits (ASICs) are playing an important role in the semiconductor products market than ever. However, the reverse engineering (RE) attacks are inducing much more severe threats to hardware design intellectual property (IP) \cite{qu2007intellectual}\cite{rostami2014primer} and great challenges for information security. Reverse engineering originally arisen from the analysis of hardware for commercial or military advantage \cite{chikofsky1990reverse}\cite{torrance2011state}\cite{guin2014counterfeit}. Usually it attempts to extract the knowledge or design information from the original creation without intellectual property permitted, and reproduce them based on the obtained information. The involved objects vary from mechanical systems, electronic devices, computer software, biological fragment, chemical samples, or any kinds of components with IPs.

The concept of intellectual property plays an important role in semiconductor industry while the recycling/remarking/redistribution constitutes a serious threat to foundries or design houses, such as Xilinx against Flextronics \cite{joel2013xilinx} in 2013. Flextronics bought Xilinx FPGA chips at a discounted rate, but remarked these devices as higher grade and sold them for elevated prices, thereby violating Xilinx IP polices through misrepresentation and exposing them to liabilities. Under the tough competition among semiconductor suppliers, some companies with reverse engineering techniques exploit the intellectual property of their competitors by incorporating the IP into their own products, without providing any credit or compensation to the IP's rightful owner, which definitely delivers quite a lot of harmfulness to research and development innovation \cite{innovation2015risk}.

The popular digital circuit watermarking and fingerprinting techniques \cite{qu2007intellectual} are passive IP protection schemes because they do not prevent RE from happening or make it more difficult. Watermark and fingerprint can be embedded into the IP to make each instance of the IP unique. When necessary, they can be revealed to show the authorship or ownership of the IP and identify the parties that misuse the IP. Although it is difficult or impossible to completely remove the watermark and fingerprint, RE attackers can still extract valuable information from the IP and reproduce the IP illegally. The existence of watermark and fingerprint in the IP can deter RE attacks, but will not increase the complexity of RE.

An active and effective approach to intellectual property protection is required, of which obfuscation is a vital solution. Circuit obfuscation seeks to modify the design and implementation of a circuit in order to make it difficult to interpret and hence increase the cost and complexity of RE attacks while the complexity to perform obfuscation for the designer is acceptable \cite{tehranipoor2015counterfeit}\cite{tehranipoor2017obfuscation}.

\subsection{Circuit Obfuscation for Anti-reverse-engineering}

Various obfuscation approaches have been proposed, of which key insertion based \cite{roy2008epic}\cite{rajendran2012security} and replacement based \cite{rajendran2013security}\cite{liu2014embeded} are two popular trends in state-of-art. Key insertion based approaches obfuscate combinational circuits by inserting some additional XOR/XNOR key-gates. Carefully selecting insertion positions can make the key extraction complexity exponential, i.e., improve the security with particular inserting procedure \cite{rajendran2012security}. Replacement based approaches replace some conventional logic gates with configurable logic units which could be programmed to perform different boolean functionalities while maintaining an identical appearance to reverse engineers. A camouflaging algorithm by judiciously selecting gates to replace with configurable CMOS cells, whose functionality can be configured by true or dummy contacts in the layout level to perform as 2-input logic gates while maintaining an identical appearance \cite{rajendran2013security}\cite{liu2014embeded}. For a reverse engineer attacker, it is very difficult to figure out the functionalities of these logic units while a popular approach is to perform attempts for many times. However, these configured true or dummy contacts still could be detected by de-layering and cross-section imaging using SEM or TEM techniques \cite{quadir2016survey} so that it still has some potential security risks. Another kind of approach is structural obfuscation which is performed by structural transformation on sequential circuits \cite{tehranipoor2017obfuscation}\cite{li2013structural}, and could even reduce the circuit area or delay compared with the un-obfuscated design occasionally. However, without a correct key or configuration, the gate replacement approach could make the IC's functionalities incorrect while the structural obfuscation only degrades the IC's performance and does not change the IC's functionalities. Therefore, the attackers could still learn some IC design details with an incorrect key even if the design is processed by structural obfuscation, which will make it be vulnerable to malicious threats such as hardware Trojans insertion. And consequently, this will make the IC less secure.

SRAM based LUT structure is adopted as configurable logic units for gate replacement while a $2^m$-to-1 multiplexer and $2^m$ configuration memory cells are utilized \cite{liu2014embeded}. Such an approach allows to dynamically configure the replaced logic gates so that the reverse engineers could not find sufficient physical/geometric information. However, adopting SRAM cell as obfuscated logic usually occupies relative high chip area overhead which are unacceptable for practical implementations. A nonvolatile reconfigurable logic is proposed as look-up-tables in \cite{zand2017energy} which is very close to our idea in this paper, however no practical applications were demonstrated in there works. A hybrid design scheme was proposed for reverse-engineering prevention by exploiting emerging spin-transfer-torque devices \cite{winograd2016hybrid}, which is very similar to our proposed methodologies in this work. However, the work in \cite{winograd2016hybrid} did not take any insight into circuit level evaluation so that it lacks of details for supporting the claimed improvements. Additionally the programming operations of STT-CMOS circuits require a relatively large current so that the CMOS transistors have to be enlarged to provide enough write current, i.e., the overhead of chip area occupation will be too large to be acceptable. In spintronics community, the spin-Hall-effect (SHE) is discovered for building a promising spin-orbit-toruqe (SOT) structure while its write current is largely reduced compared with conventional pure-STT structure \cite{miron2010current}\cite{miron2011perpendicular}\cite{liu2012current}\cite{liu2012spin}. Hence, the hybrid SOT-CMOS circuits are utilized in this work to realize the reconfigurable logic with a lower write current among LUTs programming operations.

\subsection{Main Contributions}

In this article, we investigate how to leverage SOT device as reconfigurable logic for efficient circuit obfuscation instead of SRAM as memory cells to circumvent the above drawbacks. Building LUTs with hybrid CMOS/SOT structures as obfuscated gates allows to \textit{improve the anti-reverse-engineering security with lower overhead} compared with existing approaches. In the next section, we briefly introduce some basics of circuit obfuscation and SOT mechanism. The third section presents the design of SOT-LUT structure, programming and sensing techniques in circuit level. Following that, CMOS/SOT hybrid design is carried out and the power, area, timing characteristics are extracted by various simulations. The SOT-LUT units are configured as several logic gates for replacing circuit units to perform circuit obfuscation. Experimental results are discussed later and concluding remarks are provided at last.

\begin{figure}
    \centering
    \includegraphics[scale=1.0]{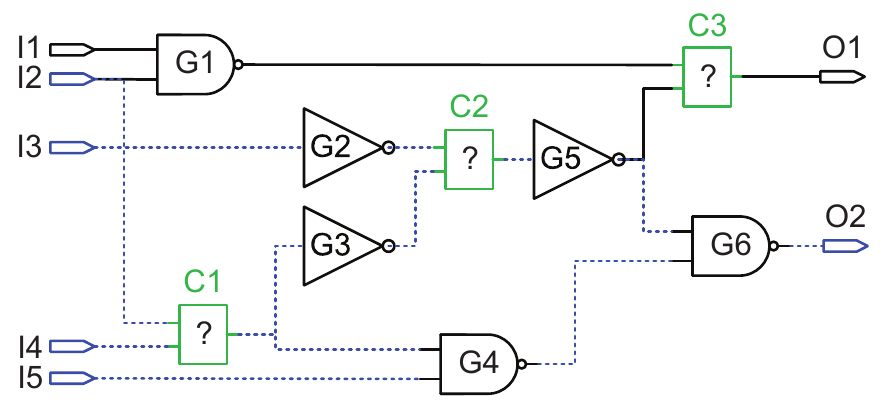}
    \caption{Circuit partition based attack example (an independent sub-circuit path is drawn in dashed lines) \cite{wang2016iscas}.}
    \label{fig:cpa}
\end{figure}

\section{Preliminary}

\subsection{Basics of Circuit Obfuscation}

The objective of chip reverse engineering is to extract the large scale layout as gate-level netlist automatically by de-packaging, de-layering, imaging and annotation \cite{quadir2016survey}. Furthermore, the gate-level netlist could be even converted to a higher-level abstraction which is very useful for the attackers to understand the system functionalities. Given an obfuscated circuit, a tricky attacker may apply various attack approaches to resolve the original netlist of IC, including testing based attack, circuit partition based attack (CPA), side channel attack, brute force attack and etc. \cite{rajendran2012security}\cite{rajendran2013security}. In this work, we will mainly evaluate our obfuscation against CPA since it is the most popular attack approach. CPA applies the divide and conquer methodology, which means an attacker can first partition the circuit to independent sub-circuits, then target each independent sub-circuit individually \cite{wang2016iscas}. An independent sub-circuit is a sub-circuits whose functionality cannot be affected by other parts, and can be tested separately from a functional IC. The attack complexity can be significantly reduced by CPA even when an obfuscation approach is resilient to all other restore attacks. As demonstrated in Fig. \ref{fig:cpa} for motivational example in \cite{rajendran2013security}\cite{wang2016iscas}, an attacker cannot resolve the functionalities of C1, C2 and C3 individually by IC testing techniques, thus has to apply brute force with a complexity of $3^3$. However, a clever attacker can first target C1 and C2 in the sub-circuit marked with dashed lines, then C3 will become an isolate unit and can be resolved in constant time. Thus the attack complexity will be $3^2$ instead of $3^3$.

\begin{figure}
    \centering
    \includegraphics[scale=1.2]{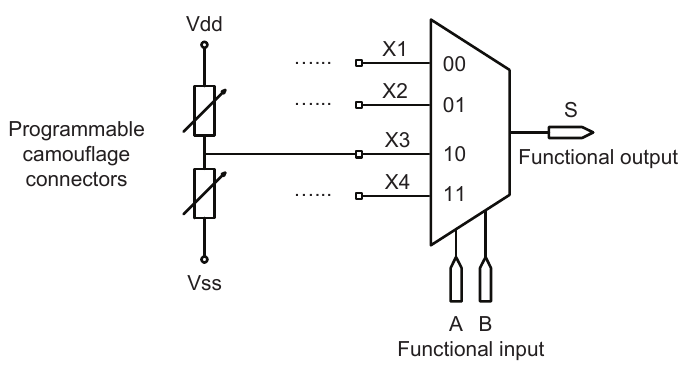}
    \caption{Configuring MUX4X1 to perform 16 possible 2-input 1-output boolean functions with camouflage connectors \cite{wang2016secure}.}
    \label{fig:mux4x1}
\end{figure}

\begin{figure*}
    \centering
    \includegraphics[scale=0.8]{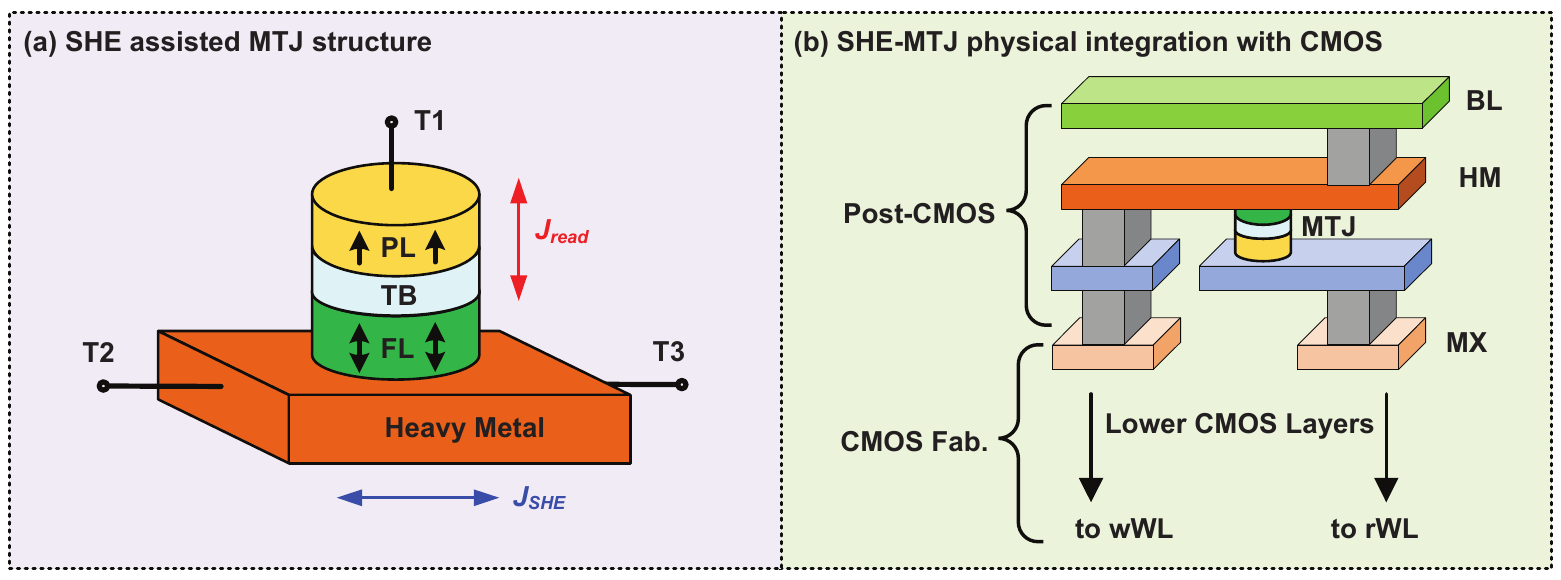}
    \caption{(a) Physical structure demonstration of spin-Hall effect assisted MTJ switching mechanism. (b) SHE-MTJ physical integration with CMOS fabrication process.}
    \label{fig:SHE}
\end{figure*}

The feasibility to perform obfuscation is arisen from the concept of reconfigurable logic by leveraging LUTs in FPGA. Multiplexers (MUXs) could be configured as different functional logics according to the different LUT-mask which is physically composed of memory bits. Besides that, configuring logic building blocks with camouflage connectors is a popular way to realize various functionalities \cite{torrance2009state}. Many approaches have been proposed to build camouflage connectors \cite{torrance2009state}. Implementing camouflage connectors is to build a particular structure in CMOS or other emerging technologies that can behave as either a connection or isolation, and appear to be physically identical under optical or electron microscopy. MUXs could be configured with camouflage connectors, with each input connected to Vdd for logic \textit{true} or Vss for logic \textit{false} by two camouflage connectors, but only one is programmed to be a connection, the other one is programmed to be an isolation. A $2^m$-to-1 MUX can be configured to perform any $2^{2^m}$ possible $m$-input boolean functions. As shown in Fig. \ref{fig:mux4x1}, a MUX4X1 is configured as 2-input logic gates. When input lines X1, X2, X3 and X4 are configured to be 1110, the MUX4X1 will perform the functionality of NAND gate \cite{wang2016secure}.

The configured structure with camouflage connectors will be realized among fabrication procedures and cannot be dynamically reconfigured after manufacturing \cite{rajendran2013security}. For the MUXs configured with memory cells, users could dynamically reconfigure their functionalities after manufacturing but these memory cells usually bring more overhead on chip area occupied \cite{liu2014embeded}. Typically the MUXs with SRAM configured structure increased the overall area by 35\%$\sim$83\% which is not easily accepted for practical implementations.

\begin{figure*}
    \centering
    \includegraphics[scale=0.8]{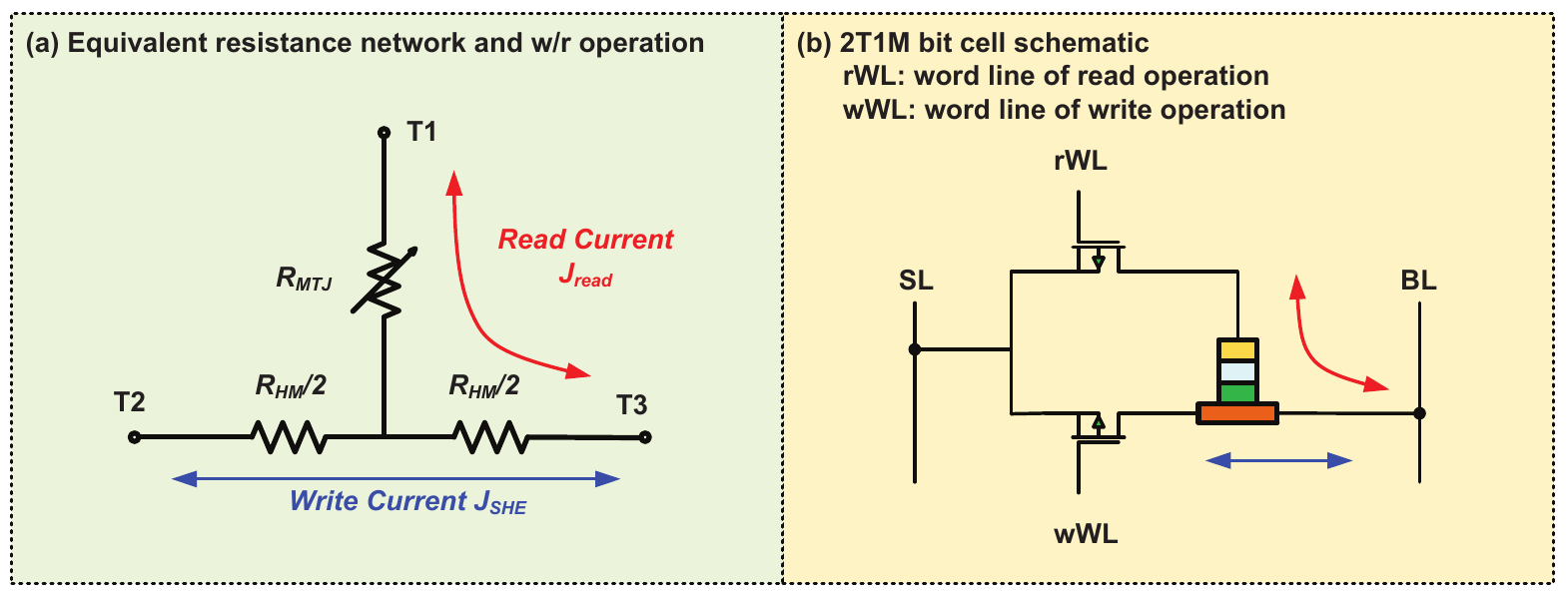}
    \caption{(a) Equivalent resistance network of SHE-MTJ write or read operations. (b) Schematic illustration of 2T1M bit cell design.}
    \label{fig:2T1M}
\end{figure*}

\subsection{Spin-Transfer and Spin-Orbit Torque Mechanism}

Spin-Transfer Torque (STT) based device has been rapidly evaluated as one of the most promising spintronics applications \cite{fongspin} with several advantages \cite{hayakawa2005current}. A typical STT device is realized by a stacked multi-layer sandwich structure that composed mainly of these layers: an oxide tunnel barrier (TB), a free magnetic layer (FL) and a pinned magnetic layer (PL) as shown in Fig. \ref{fig:SHE}(a). The magnetization direction of the free (storage) layer can be switched by an external applied magnetic field or reversed by a spin-polarized current $J_{read}$ flowing through the junction from either parallel state (P) or antiparallel state (AP) to that of the reference layer. The switching mechanism occurs in the manner of in-plane when the current density exceeds a critical density value $J_c$ that has been found lately as low as $8 \times 10^5 A/cm^2$ in CoFeB/MgO/CoFeB stack structure \cite{hayakawa2005current}. As the spin-MTJ device surface is usually small (e.g. $113nm \times 75nm$ or even less), the critical current is thereby less than one hundred $\mu A$ and could be generated by a typical CMOS current source. In addition, $J_c$ could be reduced by integrating the perpendicular magnetic anisotropy (PMA) in the free layer of the MTJ so that the stable magnetization points out-of-plane instead of in-plane \cite{ikeda2010perpendicular}\cite{peng2015origin}.

However, the popular usage of two-terminal MTJ device still poses several potential operations issues. The write path and read path always share the same current flowing branch so that it is difficult to optimize the write and read performance at the same time. Furthermore, the asymmetry property of write and read operation leads to different operation energy and delay, and thus a larger current is usually required to finish the write operation, which may lead to reliability issues. To overcome the above bottlenecks, spin-orbit interaction was recently investigated to provide an alternative write approach. Spin-orbit interaction means that the electron spin angular momentum interacts with its orbital angular momentum. In some materials, spin-orbit interaction can be strong enough to generate significant spin accumulation from an unpolarized charge current. The spin accumulation induces a torque (named spin-orbit torque, SOT) to switch the magnetization \cite{brataas2014spin}.

As shown in Fig. \ref{fig:SHE}(a), a free layer (FL) with perpendicular magnetization is sandwiched between an oxide-insulator and a nonmagnetic heavy metal (HM) strip (e.g. Pt/Ta) with strong spin-orbit interaction. The key idea of such a structure is that an in-plane charge current flowing through the heavy metal can generate the SOT for the magnetization switching, which is known as the Rashba effect \cite{miron2010current}\cite{miron2011perpendicular} or spin Hall effect (SHE) \cite{liu2012current}\cite{liu2012spin}. The SHE-induced spin current $J_{SHE}$ can be injected into FL layer which is adjacent to HM layer, resulting in a torque which could achieve the deterministic switching of perpendicular magnetization. Such a SHE-induced switching requires only on in-plane write current flowing through the heavy metal instead of through the MTJ, thus the risk of barrier breakdown could be reduced. For the in-plane magnetization, deterministic switching can be achieved by SHE current only. However, an additional magnetic field is usually required to make a deterministic switching for the perpendicular magnetization. Meanwhile, the initial SOT is more easily triggered for the case of perpendicular magnetization compared with the conventional STT since the injected spin are orthogonal to the anisotropy axis. And thus, fast write operation could be achieved with an energy efficient manner due to the required lower critical current. These advantages have been validated by the experimental demonstrations \cite{miron2010current}\cite{miron2011perpendicular}. Early research usually requires an external magnetic field to assist the switching due to the induced SHE is not enough \cite{liu2012current}\cite{liu2012spin}. But most recent research has shown that field-free magnetization reversal could be realized by spin-Hall effect and exchange bias when a perpendicularly magnetized Pt/Co/IrMn structure is introduced for ferromagnetic/anti-ferromagnetic interface \cite{lau2016spin}\cite{fukami2016magnetization}\cite{brink2016field}.

The hybrid fabrication process with MTJ and CMOS technology is shown in Fig. \ref{fig:SHE}(b) where the MTJ stack has a vertical structure similar to CMOS fabrication with low enough annealing temperatures during the process. One advantages of this fabrication is that the MTJ integration does not take much die surface except for the sensing CMOS circuits and the contacts necessary to connect the MTJs with MOS transistors. The total cost of SOT-LUT based obfuscated circuit could be lower than the SRAM-LUT based obfuscated circuit because only several additional masks are needed to integrate the SHE-MTJ at the backend process. One of the major issues for the SHE-MTJ fabrication is the height of oxide barrier, which should not be too low (e.g., $<0.7nm$) to exhibit the TMR effect and not too high (e.g., $>2.5nm$) to keep the low resistance value. A well-controlled and precise deposit process for the oxide barrier is required to avoid the mismatch variation and ensure the good SHE-MTJ sensing performance. Such hybrid CMOS/MTJ integration process has been realized by several research institutes \cite{zhao2014synchronous}\cite{everspin2006patent}. Also several magnetic memory chips have been fabricated with MTJ/CMOS integration by Hideo Ohno's group \cite{ohno2009hybrid} and Everspin Technologies \cite{mram2016company}.

\begin{figure*}
    \centering
    \includegraphics[scale=0.9,angle=0]{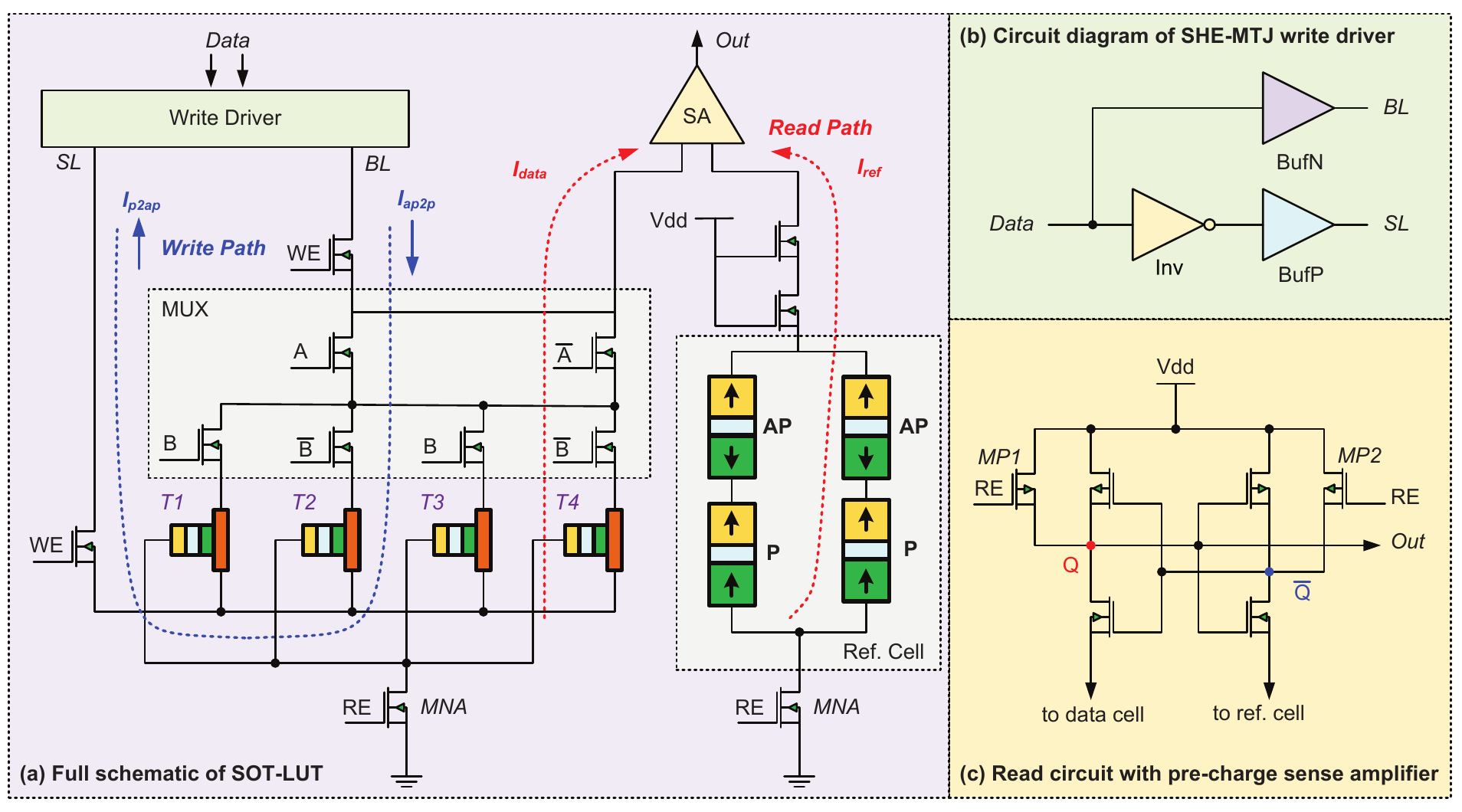}
    \caption{A 2-input SOT-LUT architecture: (a) Full schematic of SOT-LUT, (b) Circuit diagram of SHE-MTJ write driver, (c) Read circuit with pre-charge sense amplifier scheme.}
    \label{fig:lut}
\end{figure*}

The equivalent resistance network of a three-terminal SOT device is illustrated in Fig. \ref{fig:2T1M}(a) for write and read operations. Among read operation procedure, a read current $J_{read}$ is injected between T1 and T3, flowing through the equivalent resistance of stacked MTJ $R_{MTJ}$ and half of the equivalent resistance of heavy metal $R_{HM} / 2$, while the readout result is determined by high $R_{MTJ}$ in anti-parallel state or low $R_{MTJ}$ in parallel state. Among write operation procedure, a write current $J_{SHE}$ is injected between terminal T2 and terminal T3, flowing through the equivalent resistance of heavy metal $R_{HM}$, which generates a spin-orbit torque by spin-Hall effect to switch the FL layer (usually named as SHE-assisted switching). Hence, the write and read current paths have been decoupled as two independent branches so that separate optimization is available for tuning. The decoupled write and read paths of three-terminal SOT device could resolve some major disadvantages of conventional two-terminal STT device, such as high-power dissipation, selection disturbance, and larger transistors required in the CMOS writing circuit. For these reasons, SOT magnetic random access memory (SOT-MRAM) has attracted a significant amount of attention and makes it one of the best nonvolatile memory candidates \cite{wang2018high}\cite{shi2016spin}. However, the key disadvantage of SOT device is that each bit-cell requires two access transistors as shown in Fig. \ref{fig:2T1M}(b). A typical SOT memory cell requires 2T1M for one bit storage, where one access transistor is utilized as word-line for read operation and another for write operation. Such 2T1M cell results in larger bit-cell footprint so that it may not be an attractive option in high-density memory applications despite all its advantages. In this work, SOT device is adopted to build look-up-table structure while all bit-cell could share the same access transistor, which will not introduce much overhead in chip area occupied.

\section{Leveraging SOT Device as Reconfigurable Logic for Circuit Obfuscation}

\subsection{SOT Circuit Design for Reconfigurable Logic}

Leveraging spin-transfer torque or spin-orbit torque devices as reconfigurable logics have been investigated in several previous works, such as \cite{Zand2016tcas}\cite{zhao2009spin}\cite{suzuki2012area}.
Taking a 2-input SOT-LUT as an example, it includes a multiplexer (MUX), SHE-MTJ write driver and sensing circuit as shown in Fig. \ref{fig:lut}. The stored configuration bit in SHE-MTJ device is selected to perform write/read operations according to the different inputs of MUXs. A SOT-PMA (Perpendicular Magnetic Anisotropy) MTJ compact model is created with VerilogA language similar to SPINLIB \cite{spinlib} and adopted for write and read simulations in our work. The PMA-MTJ shape is $40$ $nm \times 40$ $nm$, $TMR = 150\%$, $I_{C0} \approx 60$ $\mu A$ and area product $R.A=5$ $\Omega  \cdot \mu {m^2}$. For the CMOS parts, ST-Microelectronics 45$nm$ low power design kits at $1.8V$ are adopted for write driver and sense amplifier circuits design.

Fig. \ref{fig:lut}(a) illustrates the details of such a 2-input SOT-LUT architecture. In SOT-LUT, configuration bits are stored in the MTJ cells \cite{ahari2015energy}. The SOT-LUT operations consist of two procedures: \textit{reconfiguration} and \textit{computing}. The reconfiguration procedure is to program the SHE-MTJ storing bits as the required states for the specified logic functionalities. The computing procedure is to perform the logic functionality with the input data $A$ and $B$. During the reconfiguration procedure, the $WE$ signal is active for write enable while $RE$ is inactive for read disable. The SHE-MTJs, $T1$, $T2$, $T3$ and $T4$, can be accessed either sequentially or in parallel and then programmed by applying the appropriate write current. During the computing procedure, the $WE$ signal is inactive for write disable while $RE$ is active for read enable. The multiplexer selects the corresponding MTJ based on the inputs $A$ and $B$ of the LUT. Then, the stored value in the MTJ is sensed by the pre-charge sense amplifier. The sense amplifier compares the resistance of the selected MTJ with the reference resistance and generates the appropriate voltage level at the output for a logic value.

\subsubsection{SHE-MTJ Write Driver for Reconfiguration Procedure}

\begin{figure}
    \centering
    \subfigure[The programming results of four MTJ states (waveform of ST1, ST2, ST3 and ST4 representing the MTJ state of $T1$, $T2$, $T3$ and $T4$ in Fig. \ref{fig:lut}(a)).]{
        \label{fig:writeLUT:writeMTJ}
        \includegraphics[scale=0.47]{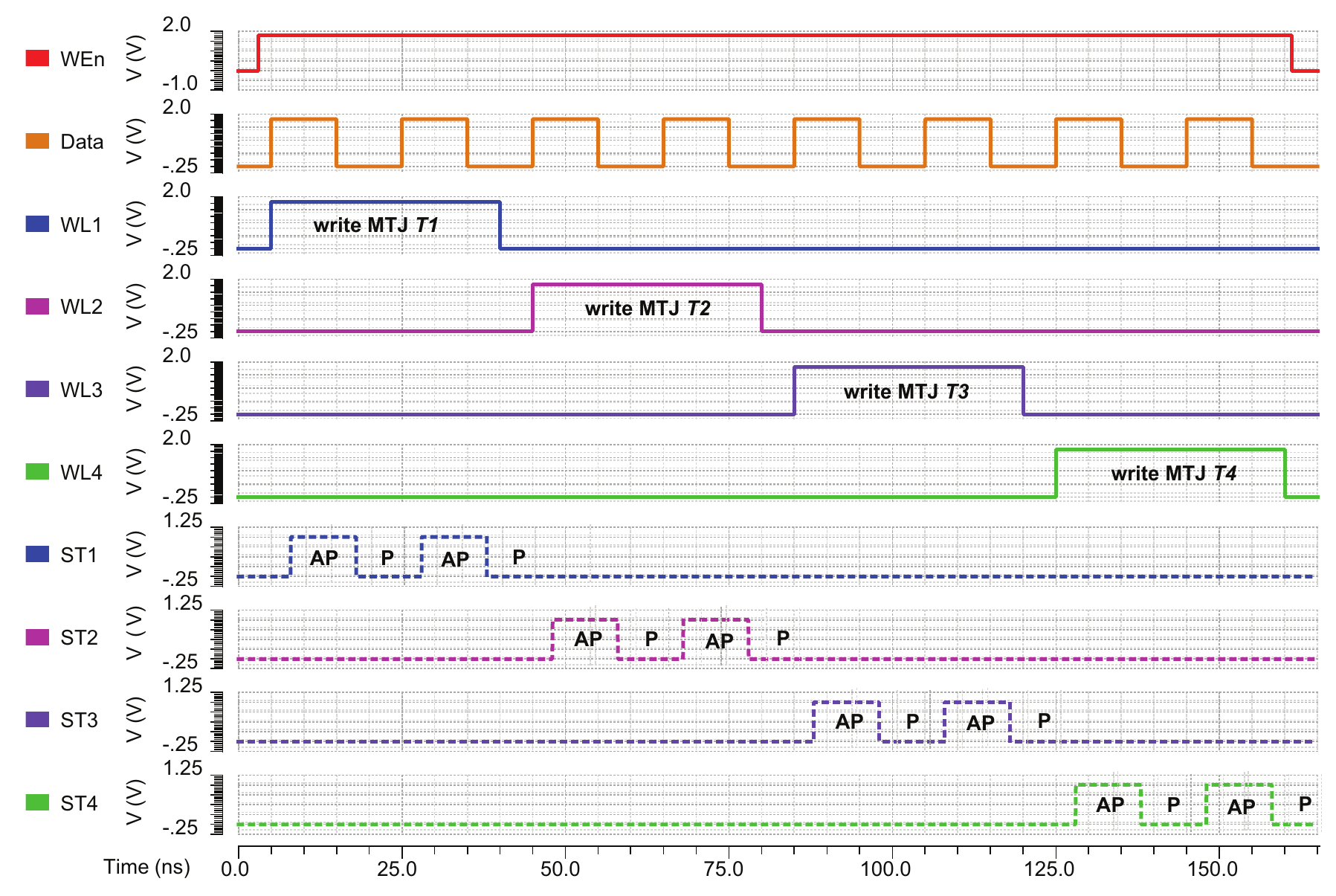}}
    \subfigure[The involved voltage drops on heavy metal of MTJ devices between bit-lines (BL) and source-lines (SL). $V_{drop}$ is about $300mV$ for writing MTJ to AP state and about $400mV$ for writing MTJ to P state.]{
        \label{fig:writeLUT:writeMTJ-SB}
        \includegraphics[scale=0.47]{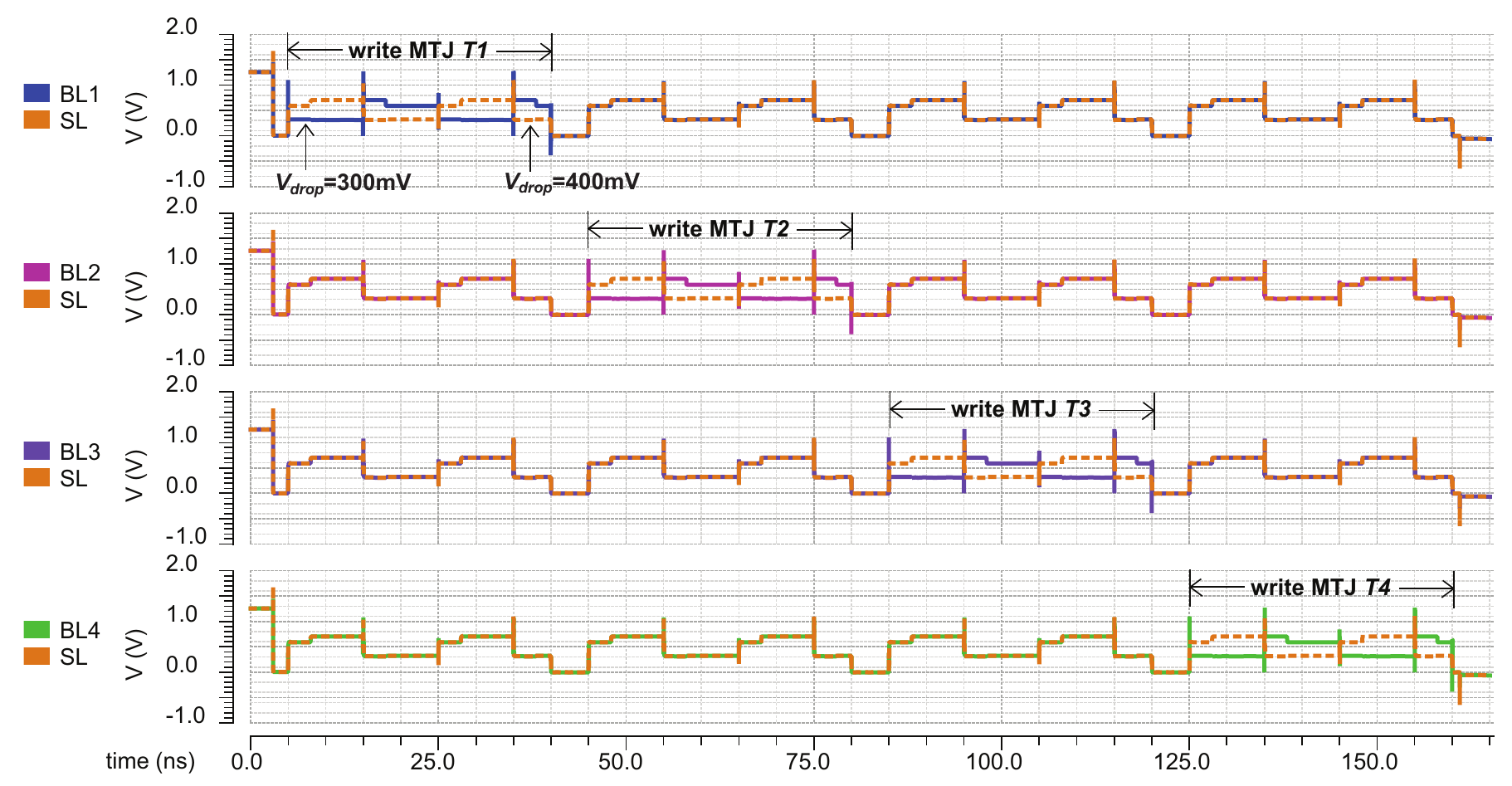}}
    \caption{Simulation results of 2-input SOT-LUT reconfiguration operations (write MTJ).}
    \label{fig:writeLUT}
\end{figure}

For the SHE-MTJ write driver, the LUT-input terminals are exploited to select the MTJ bit cell for programming according to the functionality of multiplexer. The state switching of MTJ bit cell requires a bidirectional current source, thereby a control logic unit is addressed to generate the required current passing through the selected MTJ bit according to the required programming data as shown in Fig. \ref{fig:lut}(b). Due to the write asymmetry for $P \to AP$ (\textit{slow-write}) and $AP \to P$ (\textit{fast-write}), the driving ability of write path to bit-line is enhanced by enhancing the buffering NMOS transistor size in BufN while the driving ability of write path to source-line is enhanced by enhancing the buffering PMOS transistor size in BufP \cite{bishnoi2016improving}. Benefiting from the high switching speed (lower than $1ns$) and low spin current required (less than $60 \mu A$) of SHE effect, the reconfiguration in serial will not slow the speed since it takes only some hundred nanoseconds for the reconfigurations of complex SOT-LUT with more than five inputs.

The simulation results of 2-input SOT-LUT are illustrated in Fig. \ref{fig:writeLUT}. There are four MTJs involved, $T1$, $T2$, $T3$ and $T4$ to be configured as the desired stetes, ST1, ST2, ST3 and ST4, according to the input $A$ and $B$ of MUX network. The \textit{word-lines} WL1, WL2, WL3 and WL4 are decided by MUX network and to represent whether the corresponding SHE-MTJ device is selected or not. If a certain SHE-MTJ is selected, the applied voltage on heavy metal is denoted by the voltage drop between \textit{bit-line} and \textit{source-line}, where all of these SHE-MTJs share the same \textit{source-line} and their \textit{bit-lines} are connected to the MUX network. Fig. \ref{fig:writeLUT:writeMTJ-SB} provides the details of the involved voltage drops between each BL and the common-SL. For a certain selected SHE-MTJ, the applied voltage on heavy metal is large enough to turn the AP/P state into P/AP state where the configured states are shown in Fig. \ref{fig:writeLUT:writeMTJ}.

\begin{figure}
    \centering
    \includegraphics[scale=0.49]{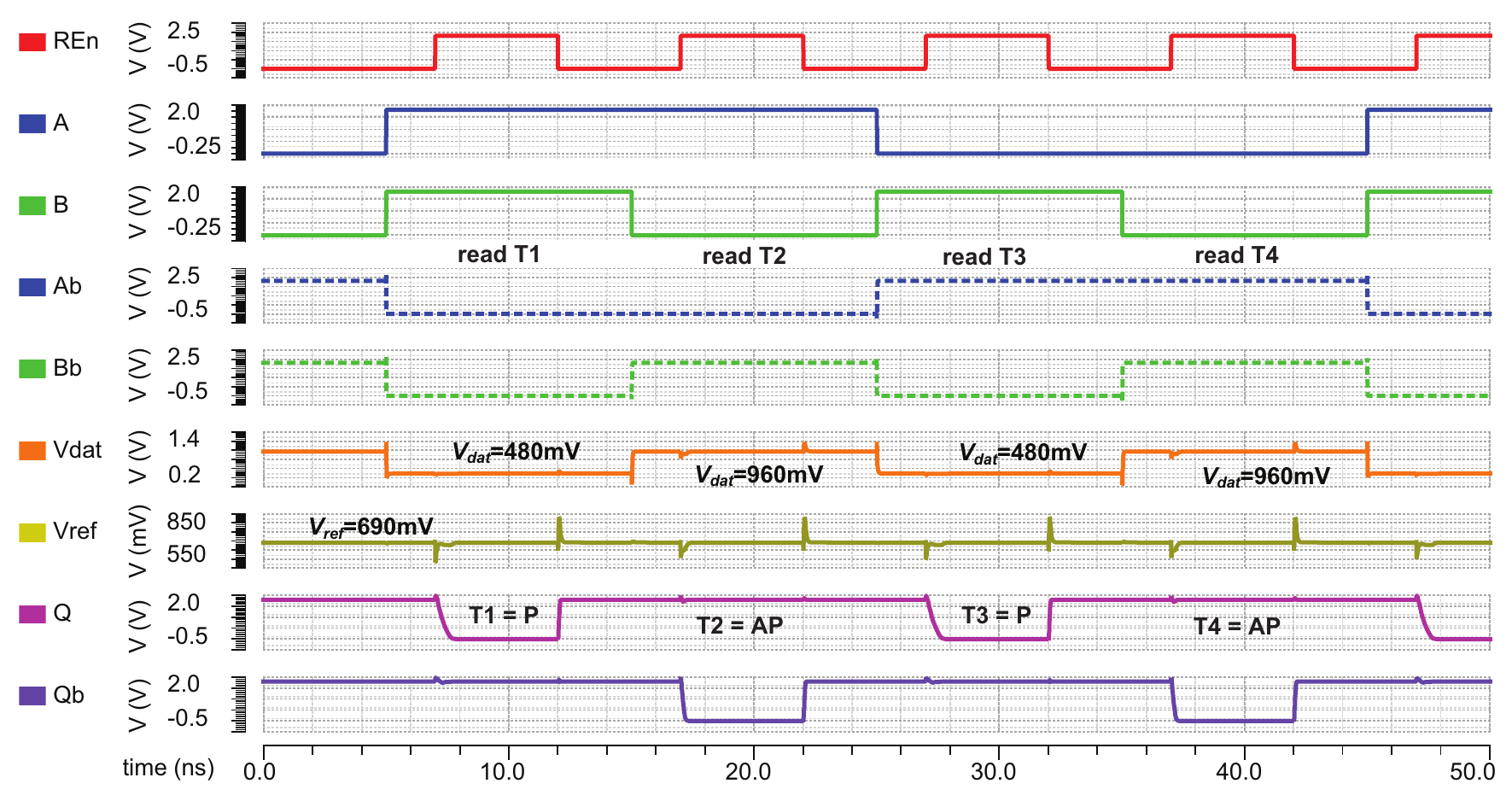}
    \caption{Simulation results of 2-input SOT-LUT pre-charge sensing operations. Wwaveform of A, B, Ab/$\rm{\overline{A}}$ and Bb/$\rm{\overline{B}}$ represents the inputs of MUX, respectively. Vdat and Vref represents the output voltage of data cell and reference cell. Vref is about $690mV$ while Vdat is about $480mV$ and $960mV$ so that the sensing margin is about $210mV$ and $270mV$ for reading P state and AP state, respectively.}
    \label{fig:readMTJ}
\end{figure}

\subsubsection{Pre-charge Sensing Amplifier During Computing}

Various sense amplifiers (SAs) have been proposed to read the state of MTJ by detecting its resistance difference. A pre-charge sense amplifier (PCSA) is adopted in our read circuit as it is capable to sense a pair of MTJs in different resistance and demonstrates very high reading speed lower than $200ps$ \cite{zhao2009high}. As shown in Fig. \ref{fig:lut}(c), the PCSA comprises two inverters, two PMOS transistors $MP1$ and $MP2$ in parallel with the pull-up PMOS of the above two inverters respectively. The pull-down NMOS transistors of the two inverters are connected to the data cell and reference cell respectively. The PCSA requires two operating phases depending on the control signal for read enable \textit{RE}. When \textit{RE} is set as low voltage, the PCSA pre-charges the mid-nodes $\rm{Q}$ and $\rm{\overline{Q}}$ of the two inverters close to \textit{Vdd} and the amplifier is kept as a metastable state. However, there is no stationary current in the circuit because the access transistor $MNA$ below is turned off by \textit{RE}.

The sensing procedure briefly begins when \textit{RE} is set as high voltage while the pre-charged voltages $\rm{Q}$ and $\rm{\overline{Q}}$ begin to discharge. Due to the resistances of the two branches are different between data cell and reference cell, the discharge speed will be different for each branch and hence the voltage of $\rm{Q}$ and $\rm{\overline{Q}}$ will be different. Furthermore, the pull-down strength of the inverters is then modulated by the voltage difference between $\rm{Q}$ and $\rm{\overline{Q}}$. For example, if the data cell is with anti-parallel state, then $R_{data} > R_{ref}$ and the discharge current in data branch is smaller than reference branch. During the modulation and feedback mechanism, the voltage $\rm{\overline{Q}}$ of reference branch will be reduced faster than the voltage $\rm{Q}$ of data branch. And when the voltage $\rm{\overline{Q}}$ becomes less than the threshold switching voltages of the inverter, the voltage $\rm{Q}$ will be charged back to $Vdd$ and $\rm{\overline{Q}}$ will continue the discharge procedure down to $Vss$. Because there are never any stationary current but only charging or discharging capacitors, the power consumption can be expected nearly to zero.

The simulation results of 2-input SOT-LUT pre-charge sensing scheme are comprehensively illustrated in Fig. \ref{fig:readMTJ}. The waveform REn represents the enable signal of amplifier circuit from Fig. \ref{fig:lut}(c) for the beginning of each read operation. The input signals, A, B, Ab (denoted as $\rm{\overline{A}}$) and Bb (denoted as $\rm{\overline{B}}$), decide which stored MTJ bit is connected to the sensing amplifier. Without loss of generality, the MTJ states of T1, T2, T3 and T4 in Fig. \ref{fig:lut}(a) have been configured as P, AP, P and AP state during reconfiguration procedure, respectively.

\subsection{Circuit Obfuscation with Reconfigurable Logic}

\subsubsection{Resources Utilization Exploration}

Without loss of generality, numerous SOT-LUTs with more inputs could be constructed similar to 2-input SOT-LUT shown in Fig. \ref{fig:lut}. Typically, additional two kinds of structures are illustrated with 3-input SOT-LUTs and 4-input SOT-LUTs which requires $2^3$ MTJs and $2^4$ MTJs respectively. These LUTs could be configured to perform 3-input or 4-input NAND/NOR/XOR and many other logic functionalities. Taking such reconfigurable SOT-LUT for circuit obfuscation, the total area occupied, power consumption and performance could be compared with conventional LUTs design. Since all of the bits of SOT-LUT share the same current source and sense amplifier, the number of transistors can be reduced greatly compared with SRAM-LUT. Three kinds of configurations are compared by counting the number of utilized transistors, including MUX-only based design, SRAM-LUT based design and SOT-LUT based design.  n-Input LUTs are constructed to implement n-Input logic gates with the number $2^n$ of storage elements. For 2-input/3-input/4-input/5-input reconfiguration logic with different number of memory cells, MUX-only based designs require 6/14/30/62 transistors, respectively. For SRAM-LUT based designs, 30/62/126/254 transistors are required, respectively. And for SOT-LUT based designs, only 27/36/53/86 transistors are required. The SOT-LUTs based reconfiguration circuits could economize up to $\sim$66\% the number of utilized transistors compared with SRAM-LUTs based circuits (6T-SRAM unit for each configurable memory cell). Hence, the SOT-LUTs based circuit obfuscation is expected to achieve a significant reduction on occupied area overhead.

It is easy to understand that the power consumption of SOT-LUT structure is very low. First, zero standby consumption benefiting from the non-volatility and high data-sensing speed, the reconfigured logic blocks in standby state could be powered down completely. For the standby power of SRAM-LUTs, the leakage power for each bit SRAM is down to about several $pW$. If we assume the SRAM cells in the 2-input LUT are in the idle state, the total power will be dominated by the standby power as high as hundreds of $pW$. This decreasing of standby power is very important for LUT applications, as it operates with stored data and there are always some logic blocks in idle state to wait the active command for most of the applications. Second, the low switching current ($\sim$200$\mu A$) significantly reduces the dynamic reconfiguration power. Based on the main-stream $45nm$ low power design kits, the SOT-LUT circuit is characterized via numerous simulations for the reconfiguration procedure and the energy consumption is about tens of $pJ$. Such a low-energy programming scheme is potentially important for energy efficient LUT implementations. Although the dynamic power of SOT-LUT is still much higher than conventional SRAM-LUT due to the required MTJ write current, it is comparatively ignored according to the significantly decreasing of standby power.

The logic operation latency of the reconfigured LUTs is critical for high speed applications. Since the adopted PCSA circuit for reading the selected SOT-LUT bit cell is similar to SRAM structure, the latency to perform the configured logic functionality is very close to conventional SRAM-LUT configured logic gate. Thus, the comparison in term of area, power and speed shows that SOT-LUT promises to replace the conventional SRAM-LUT for most of the applications.

\subsubsection{Obfuscation Procedures}

\begin{table}
\setlength{\tabcolsep}{5pt}
\setlength{\extrarowheight}{1.5pt}
\caption{A $4 \times 1$ MUX-basd LUT implementation for 16 possible 2-input 1-output Boolean functions.}
\label{tab:mux4x1}
\centering
\begin{tabular}{c!{\VRule[0.8pt]}cccc!{\VRule[0.8pt]}l}
  \specialrule{0.8pt}{0pt}{0pt}
  No. & $x_1$ & $x_2$ & $x_3$ & $x_4$ & Logic Function \\
  \hline
  1 & 0 & 0 & 0 & 1 & $F = A \cdot B$ \\
  2 & 0 & 0 & 1 & 0 & $F = A \cdot \overline{B}$ \\
  3 & 0 & 0 & 1 & 1 & $F = A + 0 \cdot B = A$ \\
  4 & 0 & 1 & 0 & 0 & $F = \overline{A} \cdot B$ \\
  5 & 0 & 1 & 0 & 1 & $F = 0 \cdot A + B = B$ \\
  6 & 0 & 1 & 1 & 0 & $F = A \oplus B$ \\
  7 & 0 & 1 & 1 & 1 & $F = A + B$ \\
  8 & 1 & 0 & 0 & 0 & $F = \overline{A} \cdot \overline{B} = \overline{A + B}$ \\
  9 & 1 & 0 & 0 & 1 & $F = A \odot B$ \\
  10 & 1 & 0 & 1 & 0 & $F = \overline{B}$ \\
  11 & 1 & 0 & 1 & 1 & $F = A + \overline{B}$ \\
  12 & 1 & 1 & 0 & 0 & $F = \overline{A}$ \\
  13 & 1 & 1 & 0 & 1 & $F = \overline{A} + B$ \\
  14 & 1 & 1 & 1 & 0 & $F = \overline{A \cdot B}$ \\
  15 & 1 & 1 & 1 & 1 & $F = const\ 1$ \\
  16 & 0 & 0 & 0 & 0 & $F = const\ 0$ \\
  \specialrule{0.8pt}{0pt}{0pt}
\end{tabular}
\vspace{-2mm}
\end{table}

The proposed SOT-LUTs are exploited for circuit obfuscation by configuring them as logic functions to replace conventional gates. Take an example as shown in Fig. \ref{fig:cpa}, the logic gates {C1, C2 and C3} could be replaced with SOT-LUTs according to the configurations illustrated in Table \ref{tab:mux4x1}. Since the attackers could not obtain the physical structures of these gates by SEM or TEM techniques, they have to try $3^2$ or $3^3$ times to resolve their functionalities. More specifically, as shown in Fig. \ref{fig:mux4x1}, a $4 \times 1$ MUX-basd LUT unit has four data lines \{$x_1$, $x_2$, $x_3$, $x_4$\}, two selection bits \{$A$, $B$\}, and one output line $F$ that comes from one of the data lines determined by the selection network. The output could be expressed as
\begin{displaymath}
F = \overline{A} \cdot \overline{B} \cdot x_1
  + \overline{A} \cdot B \cdot x_2
  + A \cdot \overline{B} \cdot x_3
  + A \cdot B \cdot x_4
\end{displaymath} By assigning the specified proper values to the input data lines, the SOT-LUT could be configured for realizing any 2-input logic functions as shown in Table \ref{tab:mux4x1}.

The circuit obfuscation in this work is to assumed to thwart Circuit Partition based Attack (CPA). Our main target is to make sure the obfuscated gates cannot be divided by any independent sub-circuit to perform attack individually. Based on these assumptions, a gate classification method is proposed in \cite{wang2016secure}, which can be leveraged in an obfuscation approach against CPA. Meanwhile, a concept named Maximum-Fan-In-Cone (MFIC) is defined to represent a set of gates whose outputs will directly or indirectly feed into specify gate. The functionality of MFIC is independent of gates and signals that do not belong to it, and can be tested or observed from inputs and outputs of this MFIC. An attacker has only access to the inputs and outputs of a functional IC, thus a MFIC is the minimum unit of an independent sub-circuit. According to the definitions, camouflaged gates from the same class (gates that belong to exactly the same set of MFIC) cannot be divided by any MFIC, thus they also cannot be divided by an independent sub-circuit to attack individually.

Once the gate classification by MFIC has been finished, we will search the inner gates of each class to find the one to be obfuscated if its design overhead of path delay is minimized. Then we use the LUTs to replace 2-input, 3-input and 4-input logic gates. When obfuscating a gate with more, or less than 2 inputs, we have additional option to first restructure this gate to make the gate end with a 2-input gate, and then obfuscate the last 2-input gate. Meanwhile, we also obfuscate the relatively large functional blocks (non-leaf cells) with single-LUT structure so that the design overhead could be further decreased compared with obfuscation with several combined LUTs.  All these obfuscation options will be evaluated and demonstrated in the experimental results.

Combining all cases we can conclude that the boolean functionalities of each $m$-input LUT appears $2^{2^m}$ possibilities to an attacker. In our obfuscation approach, we locate candidate gates based on IC structure information. We only need to apply polynomial search algorithm in the graph that represents the circuit. Thus running the obfuscation algorithm can be finished in polynomial time. For more details about the RE attack methodologies, please refer to \cite{wang2016iscas}\cite{wang2016secure}.

\subsection{Security Evaluations}

\begin{table*}[htp]
\setlength{\tabcolsep}{5pt}
\setlength{\extrarowheight}{1.5pt}
\caption{Statistics of circuit parameters for ISCAS 85/89 and MCNC benchmark suites. \# PIs is the number of input pins. \# POs is the number of output pins. \# Gates is the total number of logic gates. \# Nets is the total number of nets. \# Level is the maximum depth of the circuit graph. \textit{MFICs} and \textit{innerGate} is the MFIC number and innerGate number in largest class after classification.}
\label{tab:tab2}
\centering
\begin{tabular}{c!{\VRule[0.8pt]}ccccccc!{\VRule[0.8pt]}cc}
  \specialrule{0.8pt}{0pt}{0pt}
  \multirow{2}{*}{Benchmark Name} & \multirow{2}{*}{\# PIs} & \multirow{2}{*}{\# POs} & \multirow{2}{*}{\# Gates} & \multirow{2}{*}{\# Nets} & \multirow{2}{*}{Area} & \multirow{2}{*}{Delay} & \multirow{2}{*}{\# Level} & \multicolumn{2}{c}{Largest Class} \\
  \cline{9-10}
  & & & & & & & & \textit{MFICs} & \textit{innerGate} \\ \hline
  s713 &  54 &  42  & 142 &  290 &  359.2 &  2.67  & 16 &  1  & 18 \\
  c432  & 36  & 7 &  180  & 363 &  448 &  3.94  & 23  & 5 &  41 \\
  i2 &  201  & 1 &  222 &  460 &  598.4  & 1.82 &  10 &  1  & 221 \\
  s1196  & 32  & 32  & 398 &  897 &  1051.2  & 2.94 &  17 &  6  & 22 \\
  s1238  & 32  & 32 &  445 &  999  & 1178.4  & 3.03 &  18  & 1  & 24 \\
  too\_large  & 38 &  3  & 519  & 1125  & 1340 &  4.17 &  25  & 1  & 148 \\
  c2670  & 233 &  140  & 721  & 1324 &  1704.8  & 2.74 &  18 &  1  & 133 \\
  c3540 &  50  & 22 &  861 &  1967 &  2360.8 &  4.94  & 31 &  10 &  82 \\
  t481 &  16 &  1  & 1098  & 2596 &  3324 &  2.46  & 15  & 1  & 1097 \\
  s5378  & 199 &  213 &  1151  & 2423 &  2980.8  & 2.2 &  14  & 2 &  84 \\
  s9234 &  247  & 250  & 1505 &  3105  & 4126.4 &  4.35  & 23 &  1 &  86 \\
  c7552 &  207 &  108 &  1612  & 3372 &  4617.6 &  4.8  & 28  & 1  & 111 \\
  i10  & 257  & 224  & 1904 &  4145 &  5135.2 &  5.94  & 36  & 10 &  121 \\
  c6288 &  32  & 32 &  2267  & 5220 &  6140.8  & 15.18  & 89  & 16 &  142 \\
  s13207  & 700  & 790  & 2480 &  4741  & 6831.2 &  4.23 &  26 &  19 &  137 \\
  \specialrule{0.8pt}{0pt}{0pt}
\end{tabular}
\vspace{-2mm}
\end{table*}

The designers usually care whether the obfuscated circuits are secure, or whether the original functionalities are too difficulty to resolve for the attackers. Thus, it is necessary to analyze how our obfuscation approach can be resilient to possible restore attacks to guarantee the exponential RE complexity. Security evaluations on several attack approaches are illustrated below:

\textit{IC Testing based Attack (ITA).} In this approach, the utilized LUTs are built to hide the information of configured functionalities. To make it impossible for an attacker to resolve the functionality of a configured LUT by justification and sensitization with IC testing principles, we ensure either its input cannot be justified from input pins, or its output cannot be sensitized to output pins. For each LUT that replaces a gate connected with output pin, we ensure at least one of its inputs cannot be justified from the input pins. And for each LUT that replaces an inner gate, its output cannot be sensitized to any output pin because the gates connected to output pin of its MFICs have been obfuscated with LUTs. More details could be found in \cite{rajendran2012security}\cite{rajendran2013security}.

\textit{Circuit Partition based Attack (CPA).} In this approach, the gates from the same class are selected to perform obfuscation. LUTs in the same class belong to exactly the same MFICs, thus they cannot be divided by any MFIC to perform attacks individually \cite{wang2016iscas}.

\textit{Brute Force Attack (BFA).} Resilience to ITA of our approach forces an attacker to brute force search possible functionalities of LUTs, and resilience to CPA forces the attack to bind the LUTs together to apply brute force. Even for a 2-input LUT to replace 2-input logic gates, it has 16 possible functionalities, thus brute force attack complexity for the attacker will be no less than $16^N$, where $N$ is the number of inner gates obfuscated. And such exponential complexity actually means infeasible for the attacker to perform attacks in modern computers. Details of BFA method could be found in \cite{rajendran2013security}.

\textit{Side Channel Attack (SCA).} All LUTs have the same area, power and timing, regardless of the functionality that they are configured to perform. Thus an attacker cannot get any additional configuration information of LUTs by the side channels \cite{le2008overview}.

In summary, the proposed obfuscation scheme is resilient to possible restore attacks (including ITA, BFA, CPA, SCA), thus guarantees exponential attack complexity. Moreover, the design complexity is polynomial thus the obfuscation process can be finished in polynomial time. Actually most of the state-of-art obfuscation approaches could have two vulnerabilities: (1) they are vulnerable at least one restore attacks, resulting in a low RE complexity for an attacker; (2) the complexities to perform obfuscation in \cite{rajendran2012security}\cite{rajendran2013security} are exponential. They locate candidate gates to obfuscate by exhaustively enumerating related input patterns to see whether there exists one input pattern that simultaneously justifies the inputs and sensitizes the outputs. However, the enumeration has an exponential complexity of $2^m$, where $m$ is the number of inputs related to this gate. The elaborated obfuscation approach for fixing both vulnerabilities and detailed attack complexity analysis could be found in \cite{wang2016secure}.

\section{Experimental Results}

\subsection{Experimental Setup}

The proposed obfuscation approach with SOT-LUT circuit is validated on standard ISCAS 85/89 and MCNC benchmark suites. The obfuscation algorithm is implemented by JAVA language to perform gate classification and identify candidate gate for replacement with LUTs. The open-source ABC program is leveraged for logic synthesis \cite{brayton2010abc} and Oklahoma Stage University standard cell library based on the TSMC 0.35$\mu m$ PDK are used for overheads measurement.

The MUXs with camouflage connectors are adopted as the baseline for evaluating LUTs for obfuscation. In OSU standard cell library, only MUX2X1 is provided while we may use three MUX2X1 stacked as a MUX4X1. However, it will result in significant design overhead especially when the total number of MUX4X1 configured for obfuscation is relatively large compared with the original design. Based on the TSMC 0.35$\mu m$ PDK, a transmission gate based MUX4X1 is implemented to comprehensively configure the required 16 functionalities. Each MUX4X1 consists of 2 inverters and 8 transmission gate which is made by parallel combination of nMOS and pMOS transistors. The adopted inverter is same as the inverter in OSU standard library, and the transistors size in transmission gate is also same as the inverter, that is, channel length is 0.4$\mu m$, channel width is 2$\mu m$ and 4$\mu m$ for nMOS and pMOS respectively. Even though additional approaches could be made to optimize the MUX4X1 design, we adopt such typical design as a reference without the loss of generality.

Two kinds of LUTs are evaluated for obfuscation: SRAM-LUT and SOT-LUT. The SRAM-LUTs adopt a typical 6T SRAM structure as the configuration memory cells. The write driver and read circuits of SOT-LUTs are carefully designed and optimized for lower power consumption, high read/write reliability and area efficient. Cadence Liberate \cite{liberate2015cadence} is adopted for cell characterization of the designed MUX4X1, SRAM-LUT and SOT-LUT. According to the characterization results, we extract the pin-to-pin delay, input load, fan-out, and cell area, which is built as a look-up table and represented in the OSU035.genlib file. Meanwhile, there is an issue we have to clarify that we are focused on evaluating the LUTs-type obfuscation approaches in this work. Actually, the CPA attack complexities for different obfuscation approaches (SRAM-LUTs or SOT-LUTs) are at the same scale. These obfuscation approaches essentially show a better security compared with other dummy contact approaches \cite{rajendran2013security} since the dummy contacts still could be revealed by advanced techniques.

\begin{figure}
    \centering
    \subfigure[Area overhead.]{
        \label{fig:gate16:area}
        \includegraphics[scale=0.5]{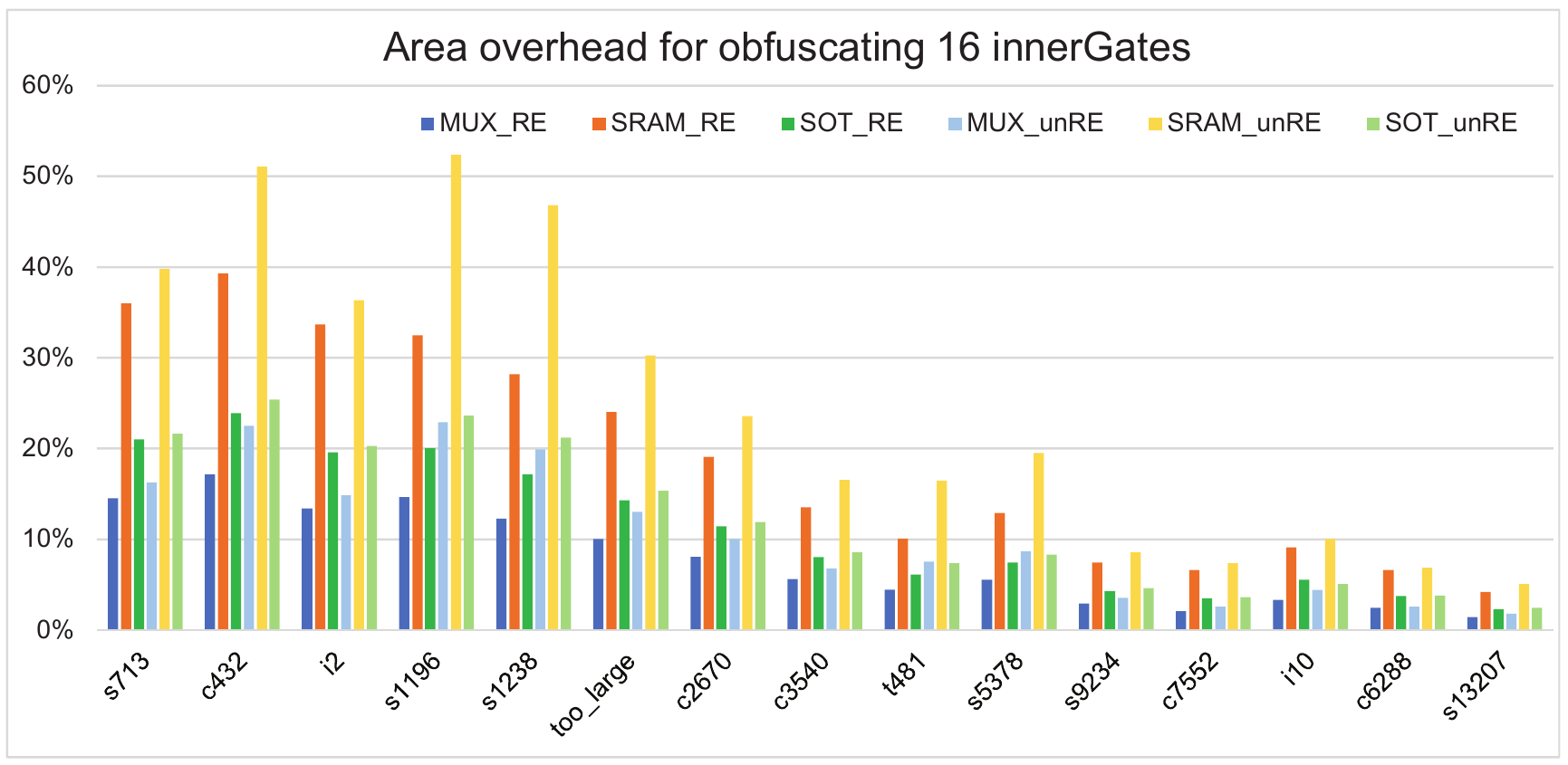}}
    \subfigure[Delay overhead.]{
        \label{fig:gate16:delay}
        \includegraphics[scale=0.5]{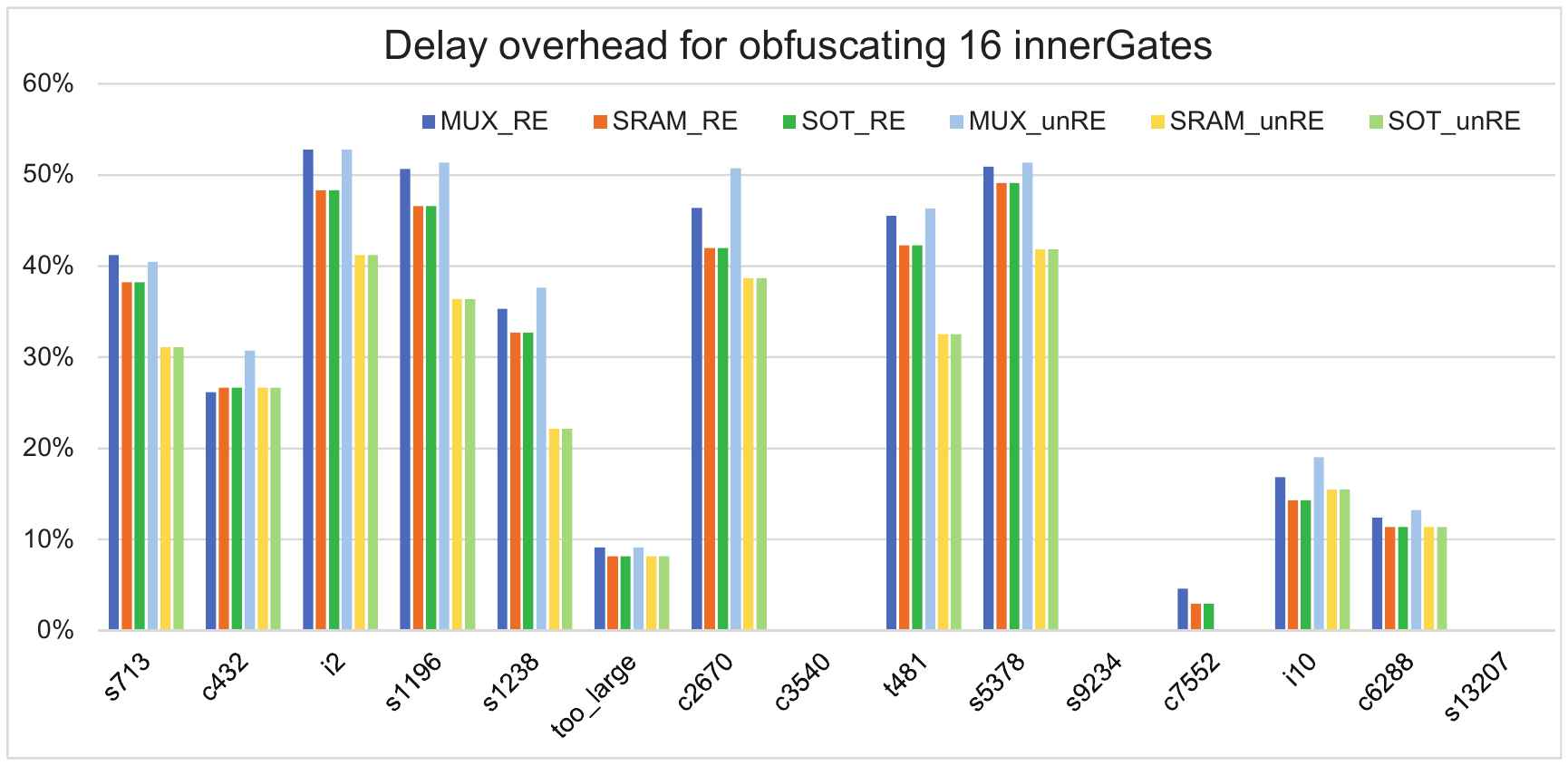}}
    \caption{Design overheads when obfuscating 16 innerGates.}
    \label{fig:gate16}
\end{figure}

\begin{figure}
    \centering
    \subfigure[Area overhead.]{
        \label{fig:gate32:area}
        \includegraphics[scale=0.5]{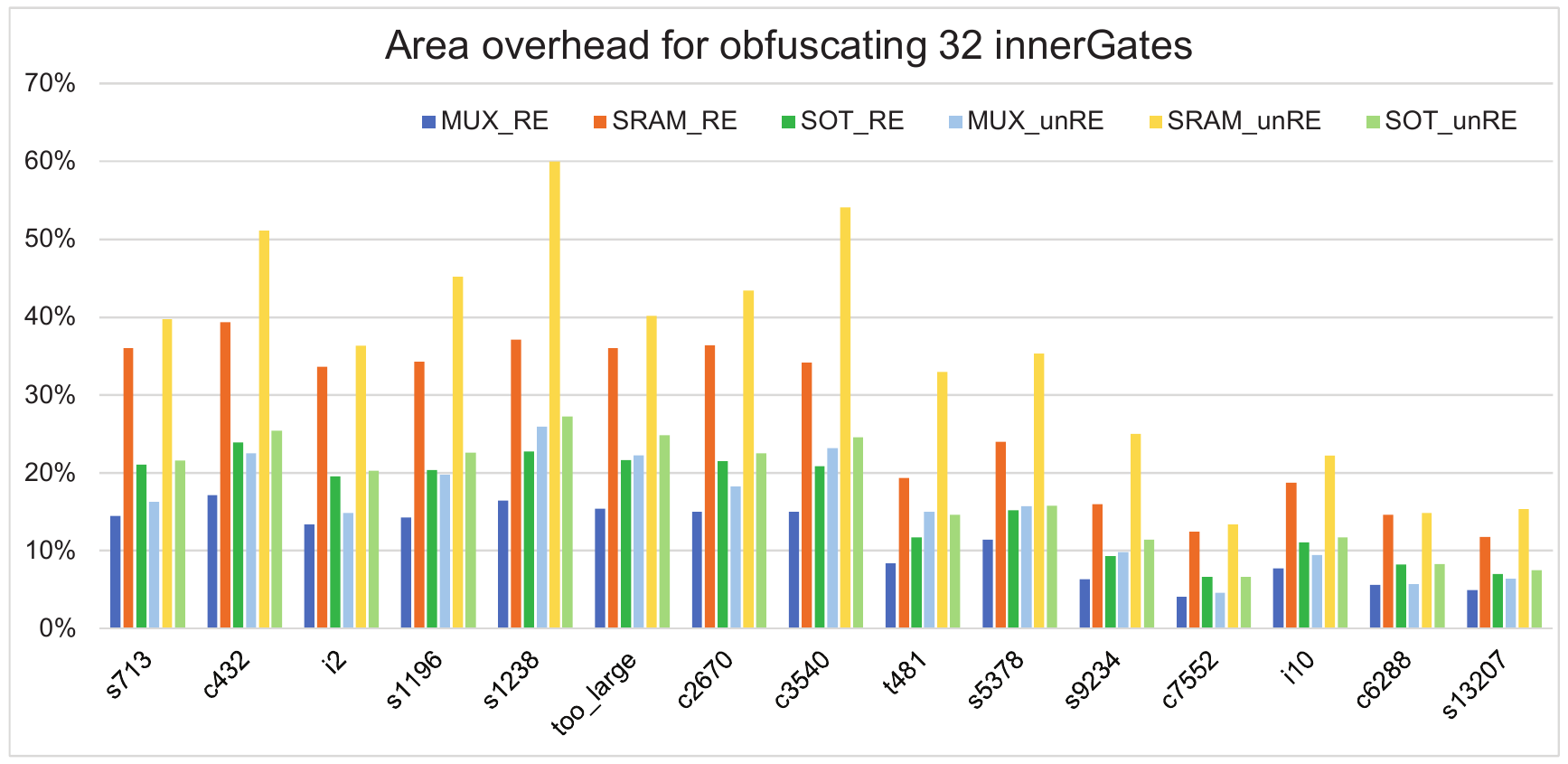}}
    \subfigure[Delay overhead.]{
        \label{fig:gate32:delay}
        \includegraphics[scale=0.5]{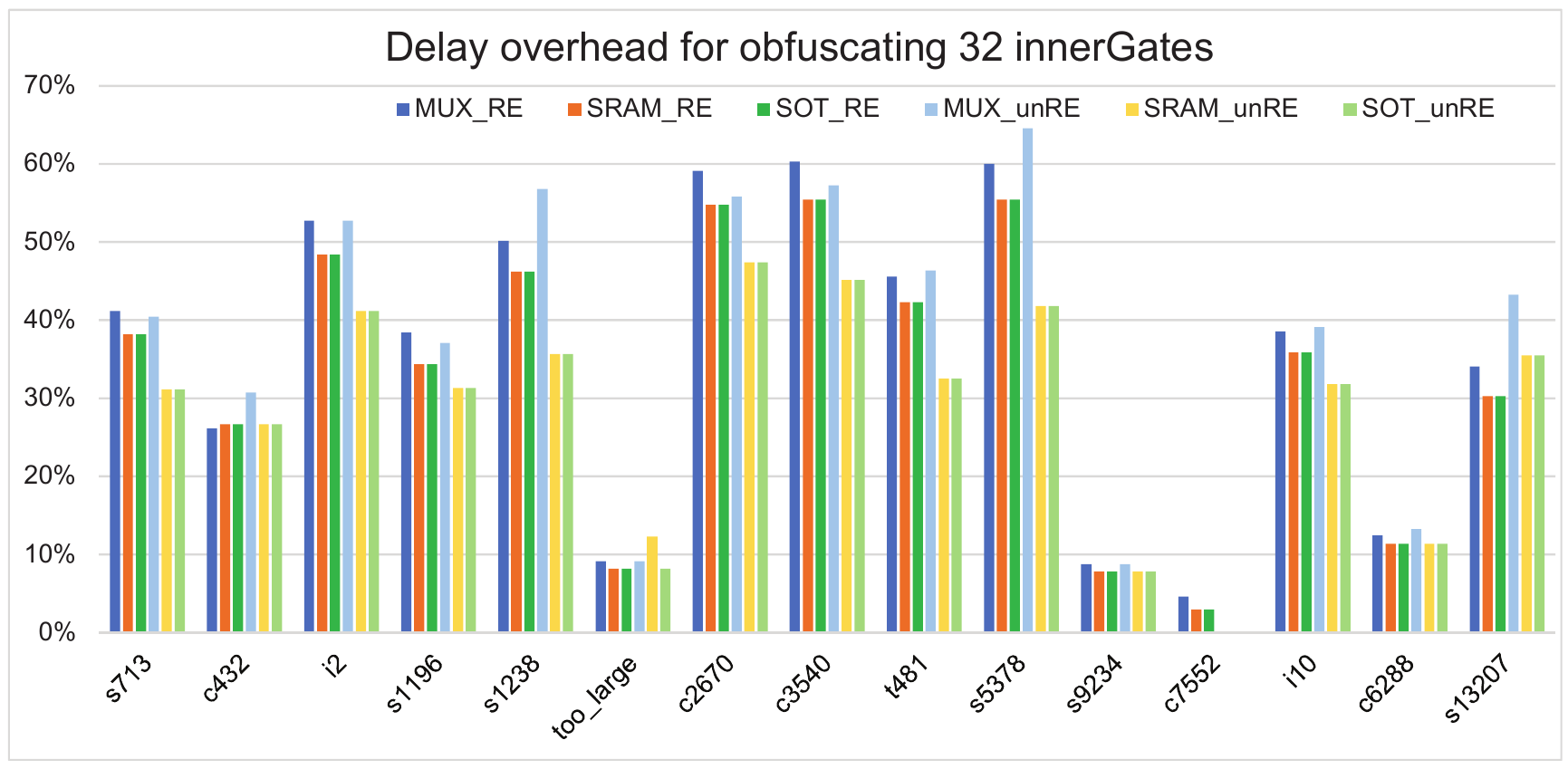}}
    \caption{Design overheads when obfuscating 32 innerGates.}
    \label{fig:gate32}
\end{figure}

\begin{table*}[htp]
\centering
\begin{threeparttable}
\setlength{\tabcolsep}{5pt}
\setlength{\extrarowheight}{1.5pt}
\caption{ Max/Min/Avg design overheads for all benchmarks with different obfuscation approaches. A composite metric $\Phi$ is defined to normalize the overall overhead of area and delay with two factors $\alpha$ and $\beta$ respectively, while $\alpha + \beta = 1$ and $\Phi = area\_overhead * \alpha + delay\_overhead * \beta$. These two factors are determined by averaging the two kinds of overheads with different significance according to the designers considerations. Taking an example, we may set $\alpha = 0.5$ and $\beta = 0.5$ if the delay overhead has the equal preference with area overhead, and consequently $\Phi = area\_overhead * 0.5 + delay\_overhead * 0.5$ by average overhead. }
\label{tab:tab3}
\centering
\begin{tabular}{c!{\VRule[0.8pt]}c!{\VRule[0.8pt]}cc!{\VRule[0.8pt]}cc!{\VRule[0.8pt]}cc!{\VRule[0.8pt]}cc!{\VRule[0.8pt]}cc!{\VRule[0.8pt]}cc}
  \specialrule{0.8pt}{0pt}{0pt}
  \multirow{2}{*}{\# $i_G$\tnote{\dag}} & \multirow{2}{*}{Overhead} & \multicolumn{2}{c!{\VRule[0.8pt]}}{MUX\_RE} & \multicolumn{2}{c!{\VRule[0.8pt]}}{SRAM\_RE} & \multicolumn{2}{c!{\VRule[0.8pt]}}{SOT\_RE} & \multicolumn{2}{c!{\VRule[0.8pt]}}{MUX\_unRE} & \multicolumn{2}{c!{\VRule[0.8pt]}}{SRAM\_unRE} & \multicolumn{2}{c}{SOT\_unRE} \\
  \cline{3-14}
  & & area & delay & area & delay & area & delay & area & delay & area & delay & area & delay \\
  \hline
  \multirow{4}{*}{16} & max & 17.14\% & 52.75\% & 39.32\% & 49.09\% & 23.89\% & 49.09\% & 22.89\% & 52.75\% & 52.39\% & 41.82\% & 25.39\% & 41.82\%  \\
  & min &  1.44\% & 0.00\% & 4.19\% & 0.00\% & 2.28\% & 0.00\% & 1.80\% & 0.00\% & 5.07\% & 0.00\% & 2.49\% & 0.00\% \\
  & average & 7.85\% & 26.12\% & 18.88\% & 24.17\% & 11.22\% & 24.17\% & 10.49\% & 26.85\% & 24.71\% & 20.36\% & 12.21\% & 20.36\% \\
  \cline{2-14}
  & \boldmath \textbf{$\Phi_{16}$}\tnote{$\ast$}  & \multicolumn{2}{c!{\VRule[0.8pt]}}{\textbf{16.99\%}} & \multicolumn{2}{c!{\VRule[0.8pt]}}{\textbf{21.53\%}} & \multicolumn{2}{c!{\VRule[0.8pt]}}{\textbf{17.69\%}} & \multicolumn{2}{c!{\VRule[0.8pt]}}{\textbf{18.67\%}} & \multicolumn{2}{c!{\VRule[0.8pt]}}{\textbf{22.54\%}} & \multicolumn{2}{c}{\textbf{16.29\%}} \\
  \hline
  \multirow{4}{*}{32} & max &  17.14\% & 60.32\% & 39.32\% & 55.47\% & 23.89\% & 55.47\% & 25.95\% & 64.55\% & 60.00\% & 47.45\% & 27.21\% & 47.45\%  \\
  & min &  4.07\% & 4.58\% & 11.74\% & 2.92\% & 6.62\% & 2.92\% & 4.61\% & 0.00\% & 13.36\% & 0.00\% & 6.66\% & 0.00\%  \\
  & average & 11.32\% & 36.07\% & 26.92\% & 33.20\% & 16.06\% & 33.20\% & 15.32\% & 37.01\% & 35.28\% & 28.76\% & 17.67\% & 28.49\% \\
  \cline{2-14}
  & \boldmath \textbf{$\Phi_{32}$}\tnote{$\ast$}  & \multicolumn{2}{c!{\VRule[0.8pt]}}{\textbf{23.70\%}} & \multicolumn{2}{c!{\VRule[0.8pt]}}{\textbf{30.06\%}} & \multicolumn{2}{c!{\VRule[0.8pt]}}{\textbf{24.63\%}} & \multicolumn{2}{c!{\VRule[0.8pt]}}{\textbf{26.17\%}} & \multicolumn{2}{c!{\VRule[0.8pt]}}{\textbf{32.02\%}} & \multicolumn{2}{c}{\textbf{23.18\%}} \\
  \hline
  \multirow{4}{*}{64} & max &   18.09\% & 85.91\% & 39.32\% & 78.64\% & 24.51\% & 78.64\% & 26.37\% & 97.27\% & 60.00\% & 78.18\% & 27.21\% & 78.18\%  \\
  & min &  10.45\% & 9.11\% & 27.17\% & 8.15\% & 15.54\% & 8.15\% & 11.28\% & 9.11\% & 28.94\% & 4.17\% & 15.81\% & 4.17\% \\
  & average & 14.34\% & 44.30\% & 33.89\% & 40.73\% & 20.29\% & 40.73\% & 19.44\% & 45.94\% & 44.62\% & 36.35\% & 22.23\% & 36.08\%  \\
  \cline{2-14}
  & \boldmath \textbf{$\Phi_{64}$}\tnote{$\ast$}  & \multicolumn{2}{c!{\VRule[0.8pt]}}{\textbf{29.32\%}} & \multicolumn{2}{c!{\VRule[0.8pt]}}{\textbf{37.31\%}} & \multicolumn{2}{c!{\VRule[0.8pt]}}{\textbf{30.51\%}} & \multicolumn{2}{c!{\VRule[0.8pt]}}{\textbf{32.69\%}} & \multicolumn{2}{c!{\VRule[0.8pt]}}{\textbf{40.49\%}} & \multicolumn{2}{c}{\textbf{29.15\%}} \\
  \specialrule{0.8pt}{0pt}{0pt}
\end{tabular}
\begin{tablenotes}
  \item[\dag] Number of the selected $innerGates$ for obfuscation.
  \item[$\ast$] Composite metric parameter is normalized by setting $\alpha = 0.5$ and $\beta = 0.5$.
\end{tablenotes}
\end{threeparttable}
\end{table*}

\subsection{Design Overheads and Attack Complexity}

The statistics of circuit parameters for the evaluated benchmarks are list in Table \ref{tab:tab2}, where column 2-8 shows the number of input pins (PIs), output pins (POs), mapped logic gates, circuit net, original area and delay before obfuscation, and the maximum depth of the circuit graph. Column 9-10 shows the MFIC number and innerGate number in largest class after performing the gate classification procedure when no nun-functional MUXs or LUTs are added. Without loss of any generality, the number of obfuscated gates is typically chosen as no more than 5\% of the total gate number for obfuscation with a reasonable overhead because this ratio is usually adopted in other state of art works\cite{rajendran2013security}. The methodologies of selecting logic gate candidates for obfuscation could refer to \cite{rajendran2013security}\cite{wang2016secure}.

\begin{figure}
    \centering
    \subfigure[Area overhead.]{
        \label{fig:gate64:area}
        \includegraphics[scale=0.5]{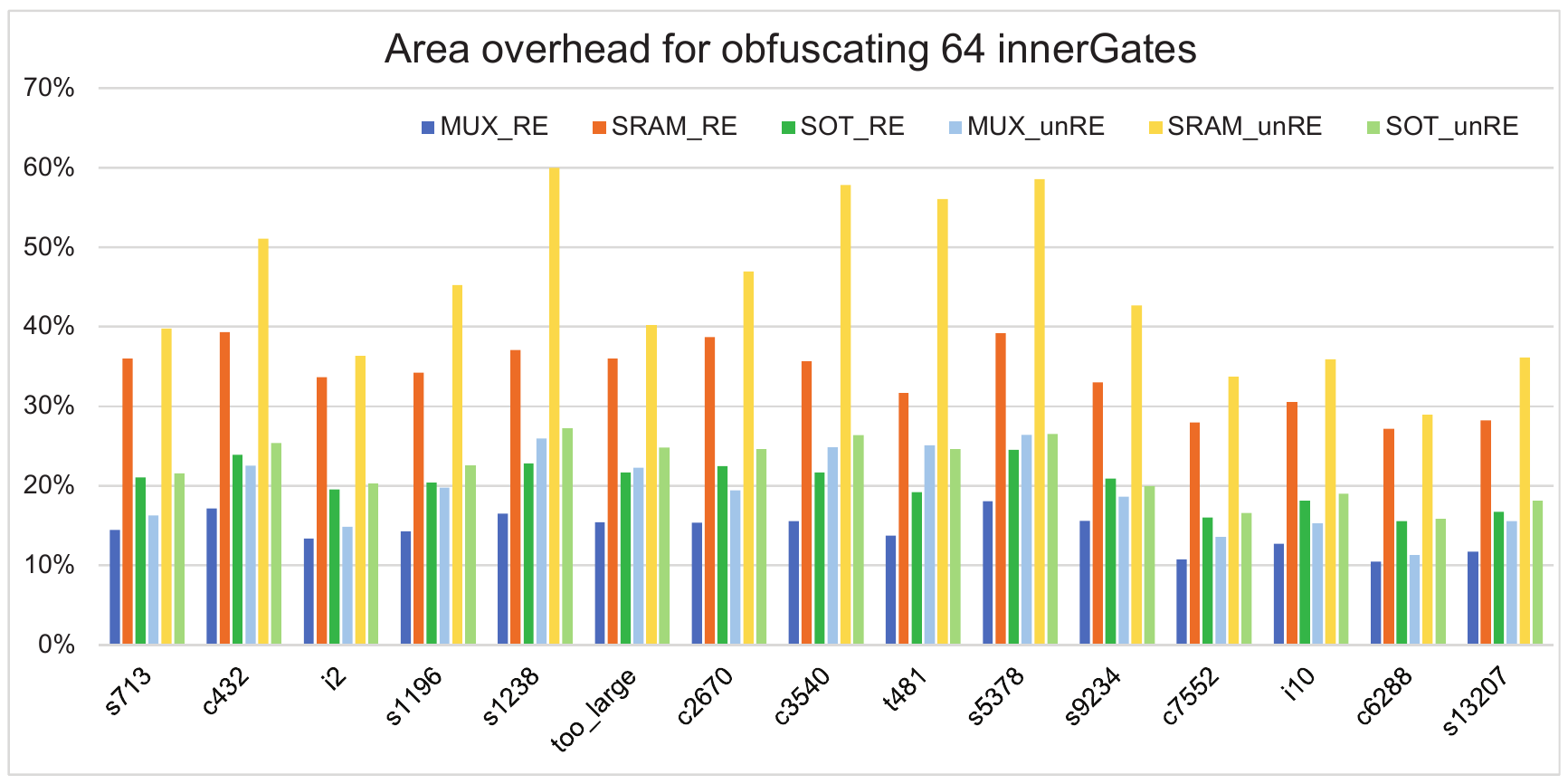}}
    \subfigure[Delay overhead.]{
        \label{fig:gate64:delay}
        \includegraphics[scale=0.5]{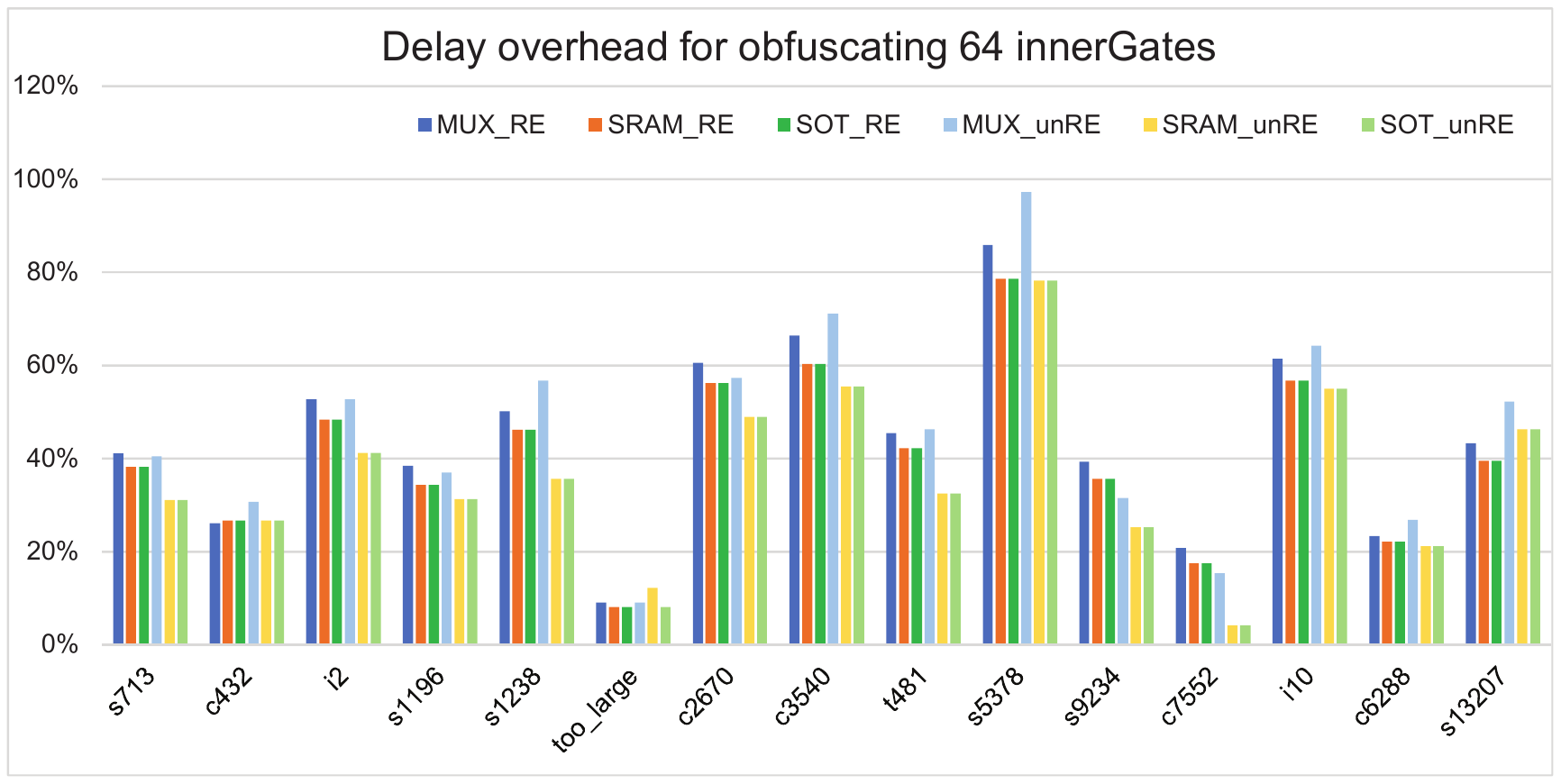}}
    \caption{Design overheads when obfuscating 64 innerGates.}
    \label{fig:gate64}
\end{figure}

The design overheads on area and delay are evaluated when obfuscating 16/32/64 innerGates, and their results are illustrated in Fig. \ref{fig:gate16}, Fig. \ref{fig:gate32} and Fig. \ref{fig:gate64}. In these cases, the attack complexity for a reverse engineer attacker can still reach $2^{64}$, $2^{128}$ and $2^{256}$, respectively, which are very high attack complexities for an attacker. Several obfuscation approaches are evaluated, including MUX\_RE, SRAM\_RE, SOT\_RE, MUX\_unRE, SRAM\_unRE and SOT\_unRE, where MUX\_RE is performing obfuscation with MUX4X1 and reconstruction strategies. SRAM\_RE is performing obfuscation with SRAM-LUT and reconstruction strategies. SOT\_RE is performing obfuscation with SOT-LUT and reconstruction strategies. MUX\_unRE is performing obfuscation with MUX4X1 but without reconstruction strategies. SRAM\_unRE is performing obfuscation with SRAM-LUT but without reconstruction strategies. SOT\_unRE is performing obfuscation with SOT-LUT but without reconstruction strategies. The utilized \textit{reconstruction} strategies attempt to realize a single MUX with multiple input by connecting several MUXs with less input. For example, a MUX4X1 could be replaced by three MUX2X1 connected.

\begin{table}
\begin{threeparttable}
\caption{Security analysis and obfuscation overhead comparison among different approaches.}
\label{tab:tab4}
\centering
\begin{tabular}{l|c|c}
  \toprule
  Approach       & Security Level & Design Overhead   \bigstrut[b]  \\
  \hline
  Dummy Contact \cite{rajendran2013security}  & Low\tnote{\dag}            &  Low          \bigstrut[t]    \\
  Pure MUX \cite{wang2016secure}      & Moderate\tnote{\ddag}         &  Moderate         \bigstrut[t] \\
  SRAM\_LUT \cite{liu2014embeded}     & High\tnote{$\ast$}           &  High        \bigstrut[t]    \\
  SOT\_LUT (this work)     & High\tnote{$\ast$}           &  Moderate            \bigstrut[t] \\
  \bottomrule
\end{tabular}
\begin{tablenotes}
  \item[\dag] Dummy contacts are visible for using SEM or TEM techniques after de-layering and cross-section imaging.
  \item[\ddag] Configured MUXs are physical connected to VDD or GND, which could be still visible if SEM or TEM is adopted.
  \item[$\ast$] Configured bits are dynamically stored, which are invisible for SEM or TEM.
\end{tablenotes}
\end{threeparttable}
\end{table}

For obfuscating 16/32/64 innerGates, the LUT-based obfuscation has more area overheads than MUX4X1-based obfuscation (as baseline) because the LUT circuit not only includes the MUXs but also other transistors or logics. However, the LUT-based obfuscation has less delay overheads compared with MUX4X1-based obfuscation because the LUTs have much more capabilities to represent complex logics with a speed-efficient manner. Hence, the tradeoff between area and speed will be very interesting for leveraging a hybrid obfuscation methodology.

The max/min/average design overheads for all benchmarks with different obfuscation approaches are illustrated in Table \ref{tab:tab3}. Noticed that the design overheads of SRAM\_RE are very close to SOT\_RE because most of the obfuscated gates are reconstructed so that 2-input LUTs are preferred to be utilized. Since the design parameters of 2-input SRAM-LUTs and SOT-LUTs are not significantly different, the total overheads between SRAM\_RE and SOT\_RE are also very similar with each other which could be validated in Table \ref{tab:tab3}. However, if the reconstruction is not deployed, the total area overheads of SRAM\_unRE is much more (about 10\%) than SOT\_unRE. Because the number of utilized MOS transistors in SOT-LUTs is less than SRAM-LUTs. Meanwhile, a composite metric $\Phi$ is defined to normalize the overall overhead of area and delay with two factors $\alpha$ and $\beta$ respectively. As shown in Table \ref{tab:tab3}, the obfuscation approach with SOT\_unRE scheme could achieve a lowest design overhead if setting $\alpha = 0.5$ and $\beta = 0.5$ equally. In summary, the adoption of SOT-LUTs for obfuscation could significantly reduce the area overhead compared with the SRAM-LUTs, and reduce the delay overhead compared with MUX-based obfuscation. Hence, the proposed SOT-LUTs are expected to be an efficient obfuscation applications for practical implementations. And hopefully, both delay and area overheads can be further reduced or controlled given a large selection pool of candidate gates in real-life circuit.

In summary, the security analysis and design overhead for several obfuscation approaches are also compared and illustrated in Table \ref{tab:tab4}. The obfuscation approach with dummy contact has a low design overhead but its security is also limited since the configured contacts could be easily revealed by SEM or TEM techniques after de-layering the chip \cite{rajendran2013security}. Using multiplexers to configure the logic functionalities could improve the attack complexity but the physically connected VDD or GND is still visible if SEM or TEM is adopted \cite{wang2016secure}. By dynamically configuring the functionalities according to SRAM\_LUT or SOT\_LUT approaches, the obfuscated logics are invisible even though the SEM or TEM techniques are exploited, which could improve the design into a higher security level. Additionally, the SOT\_LUT approach could achieve a high security level with moderate design overhead compared with SRAM\_LUT approach. Consequently the SOT\_LUT based approach will be a good choice for efficient circuit obfuscation.

\section{Conclusions}
Circuit obfuscation is a promising way to thwart reverse engineering. However, without carefully designing and obfuscation, a tricky attacker can make it really weak by performing restore attacks. We have proposed an emerging reconfigurable circuit for efficient obfuscation with lower area and delay overheads. Such circuit units exploit the magnetization states of MTJs to store the configuration memory, which is impossible to be detected using SEM or TEM techniques. The proposed SOT-LUTs obfuscation approach could eliminate the potential security risks of conventional obfuscation techniques such as MUX-based or SRAM-LUTs based approaches, and will be expected as a promising anti-reverse-engineering technique for practical IP protection applications.

\ifCLASSOPTIONcaptionsoff
  \newpage
\fi


\begin{thebibliography}{10}
\providecommand{\url}[1]{#1}
\csname url@samestyle\endcsname
\providecommand{\newblock}{\relax}
\providecommand{\bibinfo}[2]{#2}
\providecommand{\BIBentrySTDinterwordspacing}{\spaceskip=0pt\relax}
\providecommand{\BIBentryALTinterwordstretchfactor}{4}
\providecommand{\BIBentryALTinterwordspacing}{\spaceskip=\fontdimen2\font plus
\BIBentryALTinterwordstretchfactor\fontdimen3\font minus
  \fontdimen4\font\relax}
\providecommand{\BIBforeignlanguage}[2]{{%
\expandafter\ifx\csname l@#1\endcsname\relax
\typeout{** WARNING: IEEEtran.bst: No hyphenation pattern has been}%
\typeout{** loaded for the language `#1'. Using the pattern for}%
\typeout{** the default language instead.}%
\else
\language=\csname l@#1\endcsname
\fi
#2}}
\providecommand{\BIBdecl}{\relax}
\BIBdecl

\bibitem{qu2007intellectual}
G.~Qu and M.~Potkonjak, \emph{Intellectual property protection in VLSI designs:
  theory and practice}.\hskip 1em plus 0.5em minus 0.4em\relax Springer Science
  \& Business Media, 2007.

\bibitem{rostami2014primer}
M.~Rostami, F.~Koushanfar, and R.~Karri, ``A primer on hardware security:
  Models, methods, and metrics,'' \emph{Proceedings of the IEEE}, vol. 102,
  no.~8, pp. 1283--1295, 2014.

\bibitem{chikofsky1990reverse}
E.~J. Chikofsky and J.~H. Cross, ``Reverse engineering and design recovery: A
  taxonomy,'' \emph{IEEE software}, vol.~7, no.~1, pp. 13--17, 1990.

\bibitem{torrance2011state}
R.~Torrance and D.~James, ``The state-of-the-art in semiconductor reverse
  engineering,'' in \emph{Proceedings of DAC}, 2011, pp. 333--338.

\bibitem{guin2014counterfeit}
U.~Guin, K.~Huang, D.~DiMase, J.~M. Carulli, M.~Tehranipoor, and Y.~Makris,
  ``Counterfeit integrated circuits: a rising threat in the global
  semiconductor supply chain,'' \emph{Proceedings of the IEEE}, vol. 102,
  no.~8, pp. 1207--1228, 2014.

\bibitem{joel2013xilinx}
\BIBentryALTinterwordspacing
J.~Rosenblatt. (2013) {Xilinx sues Flextronics alleging fradulent chip resale,
  Bloomberg Technology}. [Online]. Available:
  \url{https://www.bloomberg.com/news/articles/2013-12-11/xilinx-sues-flextronics-alleging-fraudulent-chip-resale}
\BIBentrySTDinterwordspacing

\bibitem{innovation2015risk}
\BIBentryALTinterwordspacing
{Innovation is at risk as semiconductor equipment and materials industry loses
  up to \$4 billion annually due to IP infringement}. [Online]. Available:
  \url{http://www.semi.org/en/Press/P043775}
\BIBentrySTDinterwordspacing

\bibitem{tehranipoor2015counterfeit}
M.~M. Tehranipoor, U.~Guin, and D.~Forte, \emph{Counterfeit Integrated
  Circuits: Detection and Avoidance}.\hskip 1em plus 0.5em minus 0.4em\relax
  Springer, 2015.

\bibitem{tehranipoor2017obfuscation}
D.~Forte, S.~Bhunia, and M.~M. Tehranipoor, \emph{Hardware Protection through
  Obfuscation}.\hskip 1em plus 0.5em minus 0.4em\relax Springer, 2017.

\bibitem{roy2008epic}
J.~A. Roy, F.~Koushanfar, and I.~L. Markov, ``{EPIC}: Ending piracy of
  integrated circuits,'' in \emph{Design, Automation and Test in Europe}, 2008,
  pp. 1069--1074.

\bibitem{rajendran2012security}
J.~Rajendran, Y.~Pino, O.~Sinanoglu, and R.~Karri, ``Security analysis of logic
  obfuscation,'' in \emph{Proceedings of DAC}, 2012, pp. 83--89.

\bibitem{rajendran2013security}
J.~Rajendran, M.~Sam, O.~Sinanoglu, and R.~Karri, ``Security analysis of
  integrated circuit camouflaging,'' in \emph{Proceedings of the ACM SIGSAC
  conference on Computer \& communications security}, 2013, pp. 709--720.

\bibitem{liu2014embeded}
B.~Liu and B.~Wang, ``Embedded reconfigurable logic for asic design obfuscation
  against supply chain attacks,'' in \emph{Proceedings of DATE}, 2014, pp.
  1--6.

\bibitem{quadir2016survey}
S.~E. Quadir, J.~Chen, D.~Forte, N.~Asadizanjani, S.~Shahbazmohamadi, L.~Wang,
  J.~Chandy, and M.~Tehranipoor, ``A survey on chip to system reverse
  engineering,'' \emph{ACM Journal on Emerging Technologies in Computing
  Systems}, vol.~13, no.~1, p.~6, 2016.

\bibitem{li2013structural}
L.~Li and H.~Zhou, ``Structural transformation for best-possible obfuscation of
  sequential circuits,'' in \emph{IEEE International Symposium on
  Hardware-Oriented Security and Trust (HOST)}, 2013, pp. 55--60.

\bibitem{zand2017energy}
R.~Zand, A.~Roohi, D.~Fan, and R.~F. DeMara, ``Energy-efficient nonvolatile
  reconfigurable logic using spin hall effect-based lookup tables,'' \emph{IEEE
  Transactions on Nanotechnology}, vol.~16, no.~1, pp. 32--43, 2017.

\bibitem{winograd2016hybrid}
T.~Winograd, H.~Salmani, H.~Mahmoodi, K.~Gaj, and H.~Homayoun, ``Hybrid
  stt-cmos designs for reverse-engineering prevention,'' in \emph{Proceedings
  of DAC}, 2016, p.~88.

\bibitem{miron2010current}
I.~M. Miron, G.~Gaudin, S.~Auffret, B.~Rodmacq, A.~Schuhl, S.~Pizzini,
  J.~Vogel, and P.~Gambardella, ``Current-driven spin torque induced by the
  rashba effect in a ferromagnetic metal layer,'' \emph{Nature materials},
  vol.~9, no.~3, pp. 230--234, 2010.

\bibitem{miron2011perpendicular}
I.~M. Miron, K.~Garello, G.~Gaudin, P.-J. Zermatten, M.~V. Costache,
  S.~Auffret, S.~Bandiera, B.~Rodmacq, A.~Schuhl, and P.~Gambardella,
  ``Perpendicular switching of a single ferromagnetic layer induced by in-plane
  current injection,'' \emph{Nature}, vol. 476, no. 7359, pp. 189--193, 2011.

\bibitem{liu2012current}
L.~Liu, O.~Lee, T.~Gudmundsen, D.~Ralph, and R.~Buhrman, ``Current-induced
  switching of perpendicularly magnetized magnetic layers using spin torque
  from the spin hall effect,'' \emph{Physical review letters}, vol. 109, no.~9,
  p. 096602, 2012.

\bibitem{liu2012spin}
L.~Liu, C.-F. Pai, Y.~Li, H.~Tseng, D.~Ralph, and R.~Buhrman, ``Spin-torque
  switching with the giant spin hall effect of tantalum,'' \emph{Science}, vol.
  336, no. 6081, pp. 555--558, 2012.

\bibitem{wang2016iscas}
X.~Wang, Q.~Zhou, Y.~Cai, and G.~Qu, ``Is the secure {IC} camouflaging really
  secure?'' in \emph{Proceedings of IEEE ISCAS}, 2016, pp. 1710--1713.

\bibitem{wang2016secure}
X.~Wang, X.~Jia, Q.~Zhou, Y.~Cai, J.~Yang, M.~Gao, and G.~Qu, ``Secure and
  low-overhead circuit obfuscation technique with multiplexers,'' in
  \emph{Proceedings of ACM GLSVLSI}, 2016, pp. 133--136.

\bibitem{torrance2009state}
R.~Torrance and D.~James, ``The state-of-the-art in {IC} reverse engineering,''
  in \emph{Proceedings of Cryptographic Hardware and Embedded Systems}.\hskip
  1em plus 0.5em minus 0.4em\relax Springer, 2009, pp. 363--381.

\bibitem{fongspin}
X.~Fong, Y.~Kim, R.~Venkatesan, S.~Choday, A.~Raghunathan, and K.~Roy,
  ``Spin-transfer torque memories: Devices, circuits, and systems,''
  \emph{Proceedings of IEEE}, 2016.

\bibitem{hayakawa2005current}
J.~Hayakawa, S.~Ikeda, Y.~M. Lee, R.~Sasaki, T.~Meguro, F.~Matsukura,
  H.~Takahashi, and H.~Ohno, ``Current-driven magnetization switching in
  {CoFeB/MgO/CoFeB} magnetic tunnel junctions,'' \emph{Japanese Journal of
  Applied Physics}, vol.~44, no.~9L, p. L1267, 2005.

\bibitem{ikeda2010perpendicular}
S.~Ikeda, K.~Miura, H.~Yamamoto, K.~Mizunuma, H.~Gan, M.~Endo, S.~Kanai,
  J.~Hayakawa, F.~Matsukura, and H.~Ohno, ``A perpendicular-anisotropy
  {CoFeB--MgO} magnetic tunnel junction,'' \emph{Nature Materials}, vol.~9,
  no.~9, pp. 721--724, 2010.

\bibitem{peng2015origin}
S.~Peng, M.~Wang, H.~Yang, L.~Zeng, J.~Nan, J.~Zhou, Y.~Zhang, A.~Hallal,
  M.~Chshiev, K.~L. Wang \emph{et~al.}, ``Origin of interfacial perpendicular
  magnetic anisotropy in {MgO/CoFe/metallic} capping layer structures,''
  \emph{Scientific reports}, vol.~5, 2015.

\bibitem{brataas2014spin}
A.~Brataas and K.~M. Hals, ``Spin-orbit torques in action,'' \emph{Nature
  nanotechnology}, vol.~9, no.~2, pp. 86--88, 2014.

\bibitem{lau2016spin}
Y.-C. Lau, D.~Betto, K.~Rode, J.~Coey, and P.~Stamenov, ``Spin--orbit torque
  switching without an external field using interlayer exchange coupling,''
  \emph{Nature nanotechnology}, vol.~11, no.~9, pp. 758--762, 2016.

\bibitem{fukami2016magnetization}
S.~Fukami, C.~Zhang, S.~DuttaGupta, and H.~Ohno, ``Magnetization switching by
  spin-orbit torque in an antiferromagnet/ferromagnet bilayer system,''
  \emph{Nature Materials}, vol.~15, no.~2, pp. 535--541, 2016.

\bibitem{brink2016field}
A.~v.~d. Brink, G.~Vermijs, A.~Solignac, J.~Koo, J.~T. Kohlhepp, H.~J. Swagten,
  and B.~Koopmans, ``Field-free magnetization reversal by spin-hall effect and
  exchange bias,'' \emph{Nature Communications}, vol.~7, no.~4, 2016.

\bibitem{zhao2014synchronous}
W.~Zhao, M.~Moreau, E.~Deng, Y.~Zhang, J.-M. Portal, J.-O. Klein, M.~Bocquet,
  H.~Aziza, D.~Deleruyelle, C.~Muller \emph{et~al.}, ``Synchronous non-volatile
  logic gate design based on resistive switching memories,'' \emph{IEEE
  Transactions on Circuits and Systems I: Regular Papers}, vol.~61, no.~2, pp.
  443--454, 2014.

\bibitem{everspin2006patent}
K.~H. Smith, B.~R. Butcher, G.~W. Grynkewich, S.~V. Pietambaram, and N.~D.
  Rizzo, ``Magnetic tunnel junction memory and method with etch-stop layer,''
  US Patent 7445943, 2006.

\bibitem{ohno2009hybrid}
H.~Ohno, ``A hybrid {CMOS}/magnetic tunnel junction approach for nonvolatile
  integrated circuits,'' in \emph{IEEE Symposium on VLSI Technology}, 2009, pp.
  122--123.

\bibitem{mram2016company}
\BIBentryALTinterwordspacing
Everspin Technologies Inc. [Online]. Available: \url{http://www.everspin.com/}
\BIBentrySTDinterwordspacing

\bibitem{wang2018high}
Z.~Wang, L.~Zhang, M.~Wang, Z.~Wang, D.~Zhu, Y.~Zhang, and W.~Zhao,
  ``High-density nand-like spin transfer torque memory with spin orbit torque
  erase operation,'' \emph{IEEE Electron Device Letters}, 2018.

\bibitem{shi2016spin}
Q.~Shi, Z.~Wang, Y.~Gao, L.~Chang, W.~Kang, Y.~Zhang, and W.~Zhao, ``A spin
  {Hall} effect-based multi-level cell for {MRAM},'' in \emph{Proceedings of
  IEEE/ACM NANOARCH}, 2016, pp. 143--144.

\bibitem{Zand2016tcas}
R.~Zand, A.~Roohi, S.~Salehi, and R.~F. DeMara, ``Scalable adaptive spintronic
  reconfigurable logic using area-matched {MTJ} design,'' \emph{IEEE
  Transcations on Circuits and Systems II: Express Briefs}, vol.~63, no.~7, pp.
  678--682, 2016.

\bibitem{zhao2009spin}
W.~Zhao, E.~Belhaire, C.~Chappert, and P.~Mazoyer, ``Spin transfer torque
  {(STT)-MRAM}--based runtime reconfiguration {FPGA} circuit,'' \emph{ACM
  Transactions on Embedded Computing Systems}, vol.~9, no.~2, p.~14, 2009.

\bibitem{suzuki2012area}
D.~Suzuki, M.~Natsui, and T.~Hanyu, ``Area-efficient {LUT} circuit design based
  on asymmetry of {MTJ}'s current switching for a nonvolatile {FPGA},'' in
  \emph{Proceedings of IEEE MWSCAS}, 2012, pp. 334--337.

\bibitem{spinlib}
\BIBentryALTinterwordspacing
W.~Zhao. {{SPINLIB}: Spintronics nanodevice {SPICE}-compatible compact model
  libarary}. [Online]. Available:
  \url{http://www.ief.u-psud.fr/~zhao/spinlib.html}
\BIBentrySTDinterwordspacing

\bibitem{ahari2015energy}
A.~Ahari, M.~Ebrahimi, and M.~B. Tahoori, ``Energy efficient partitioning of
  dynamic reconfigurable {MRAM-FPGAs},'' in \emph{Proceedings of IEEE FPL},
  2015, pp. 1--6.

\bibitem{bishnoi2016improving}
R.~Bishnoi, M.~Ebrahimi, F.~Oboril, and M.~Tahoori, ``Improving write
  performance for {STT-MRAM},'' \emph{IEEE Transactions on Magnetics}, 2016.

\bibitem{zhao2009high}
W.~Zhao, C.~Chappert, V.~Javerliac, and J.-P. Nozi{\`e}re, ``High speed, high
  stability and low power sensing amplifier for {MTJ/CMOS} hybrid logic
  circuits,'' \emph{IEEE Transactions on Magnetics}, vol.~45, no.~10, pp.
  3784--3787, 2009.

\bibitem{le2008overview}
T.-H. Le, C.~Canovas, and J.~Cl{\'e}diere, ``An overview of side channel
  analysis attacks,'' in \emph{Proceedings of the 2008 ACM symposium on
  Information, computer and communications security}.\hskip 1em plus 0.5em
  minus 0.4em\relax ACM, 2008, pp. 33--43.

\bibitem{brayton2010abc}
R.~Brayton and A.~Mishchenko, ``{ABC}: An academic industrial-strength
  verification tool,'' in \emph{Computer Aided Verification}.\hskip 1em plus
  0.5em minus 0.4em\relax Springer, 2010, pp. 24--40.

\bibitem{liberate2015cadence}
\BIBentryALTinterwordspacing
(May 20, 2015) {Virtuoso Liberate Characterization Solution, Cadence Design
  Systems inc.} [Online]. Available: \url{http://www.cadence.com}
\BIBentrySTDinterwordspacing

\end{thebibliography}



\vspace{-5mm}

\begin{IEEEbiography}[{\includegraphics[width=1in,height=1.25in,clip,keepaspectratio]{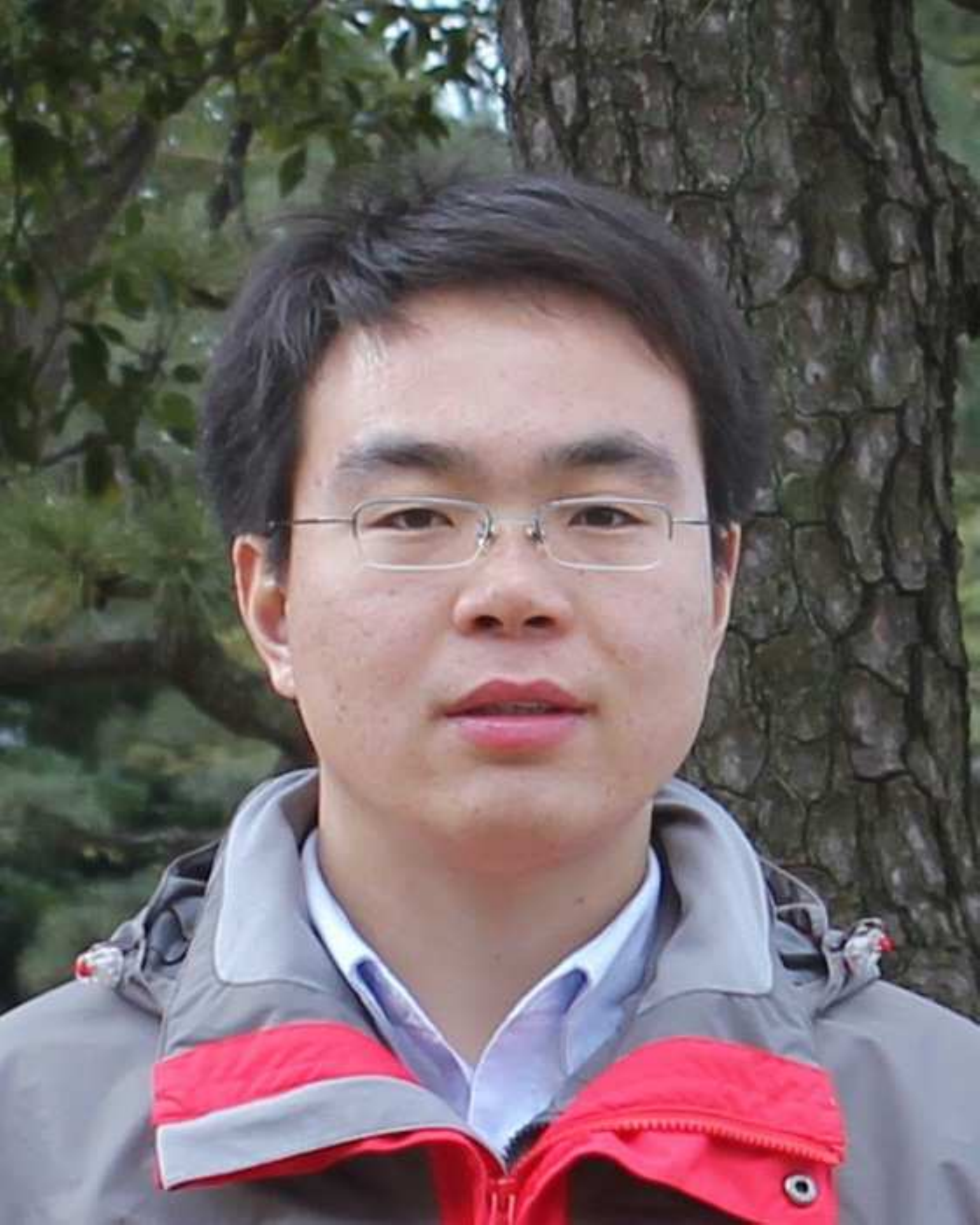}}]{Jianlei Yang}

(S'12-M'16) received the B.S. degree in microelectronics from Xidian University, Xi'an, China, in 2009, and the Ph.D. degree in computer science and technology with Tsinghua University, Beijing, China, in 2014.

He joined Beihang University, Beijing, China, in 2016, where he is currently an Associate Professor with the School of Computer Science and Engineering. From 2014 to 2016, he was a post-doctoral researcher with the Department of Electrical and Computer Engineering, University of Pittsburgh, Pittsburgh, Pennsylvania, United States. From 2013 to 2014, he was a research intern at Intel Labs China, Intel Corporation. His current research interests include STT-MRAM design and neuromorphic computing systems.

Dr. Yang was the recipient of the first place on TAU Power Grid Simulation Contest in 2011, and the second place on TAU Power Grid Transient Simulation Contest in 2012. He was a recipient of IEEE ICCD Best Paper Award in 2013, IEEE ICESS Best Paper Award in 2017, and ACM GLSVLSI Best Paper Nomination in 2015.

\end{IEEEbiography}

\vspace{-5mm}

\begin{IEEEbiography}[{\includegraphics[width=1in,height=1.25in,clip,keepaspectratio]{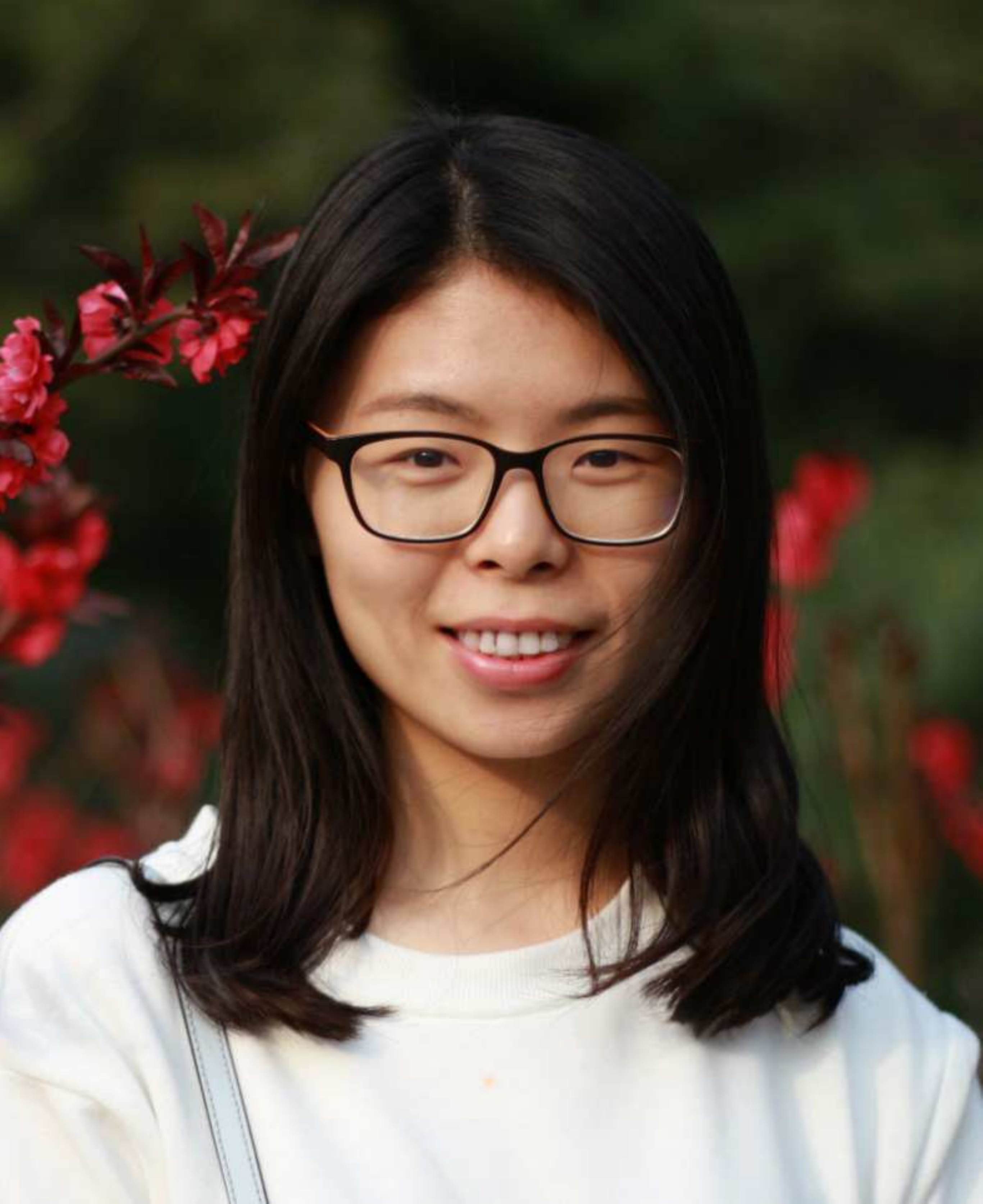}}]{Xueyan Wang}

(S'14) received the B.S. degree in computer science from Shandong University, Jinan, China, in 2013. She is currently pursuing the Ph.D. degree in computer science and technology with Tsinghua University, Beijing, China.

She is involved in research with the EDA Laboratory. From 2015 to 2016, she was a visiting student in University of Maryland, College Park, MD, USA. Her current research interests include efficient algorithms for VLSI physical design and hardware security.

\end{IEEEbiography}

\vspace{-5mm}

\begin{IEEEbiography}[{\includegraphics[width=1in,height=1.25in,clip,keepaspectratio]{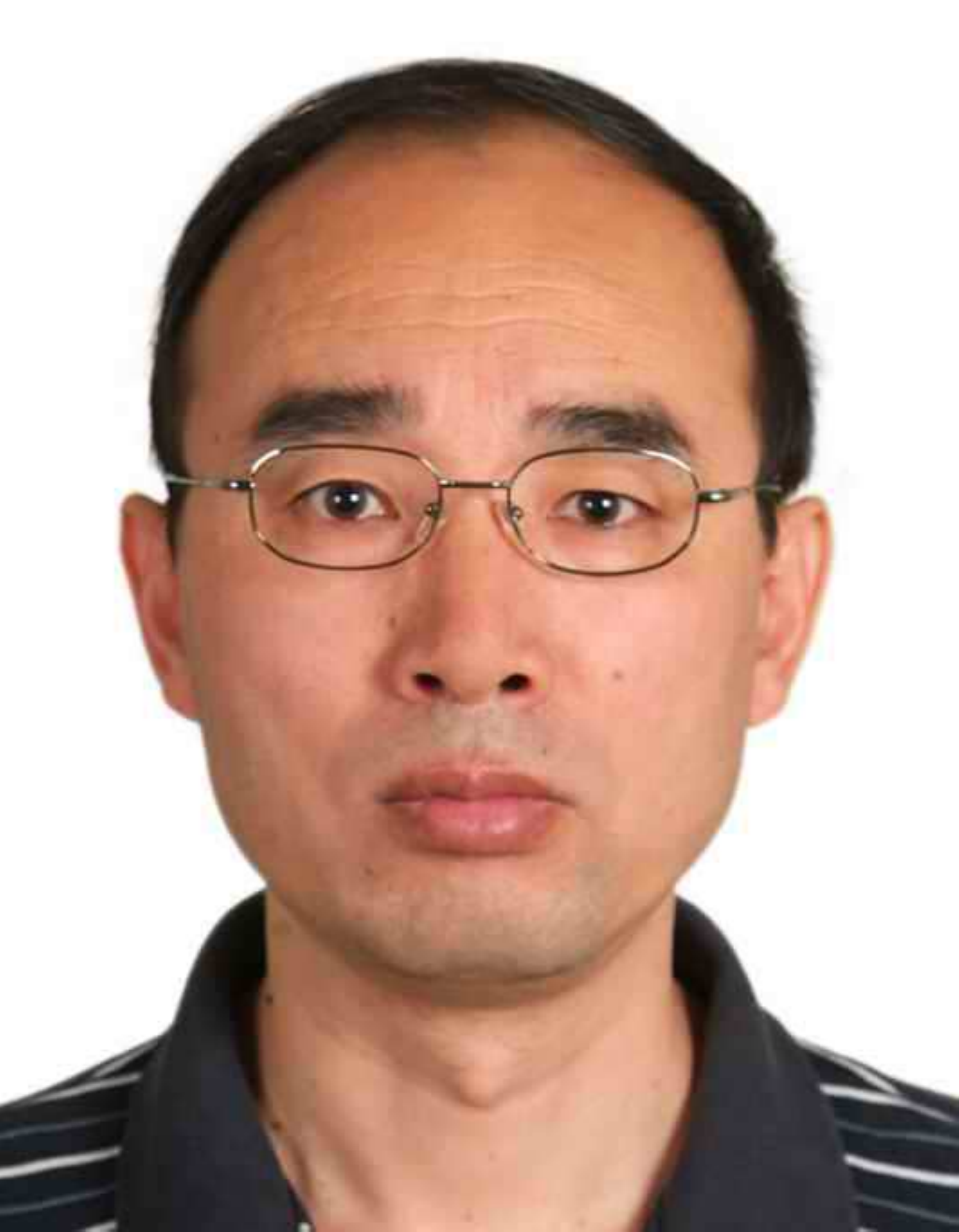}}]{Qiang Zhou}

(M'04-SM'10) received the B.S. degree in computer science and technology from the University of Science and Technology of China, Hefei, China, the M.S degree in computer science and technology from Tsinghua University, Beijing, China, and the Ph.D. degree in control theory and control engineering from the Chinese University of Mining and Technology, Beijing, in 1983, 1986 and 2002, respectively.

He has been a Professor with the Department of Computer Science and Technology, Tsinghua University. His current research interests include VLSI layout theory and algorithms.

\end{IEEEbiography}

\vspace{-5mm}

\begin{IEEEbiography}[{\includegraphics[width=1in,height=1.25in,clip,keepaspectratio]{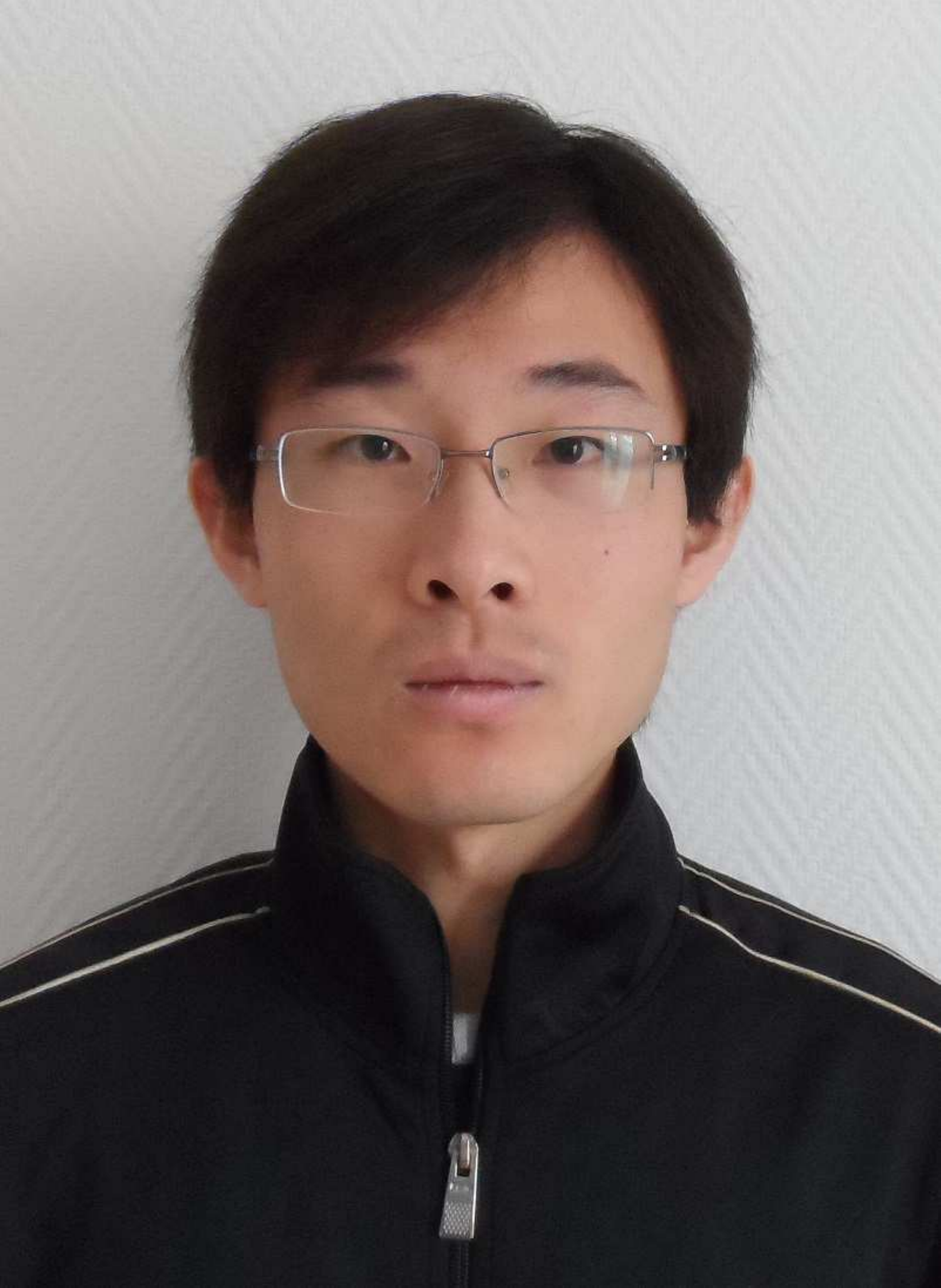}}]{Zhaohao Wang}

(S'12-M'16) received the B.S. degree in Microelectronics from Tianjin University in 2009, M.S. degree in Beihang University in 2012, and Ph.D degree in Physics from Univ. Paris-Sud, France, in 2015.

He has been an assistant Professor with the School of Electronic and Information Engineering in Beihang University since 2016. His current research interests include non-volatile nano-devices modeling, and non-volatile memory/logic circuits design.

\end{IEEEbiography}

\vspace{-5mm}

\begin{IEEEbiography}[{\includegraphics[width=1in,height=1.25in,clip,keepaspectratio]{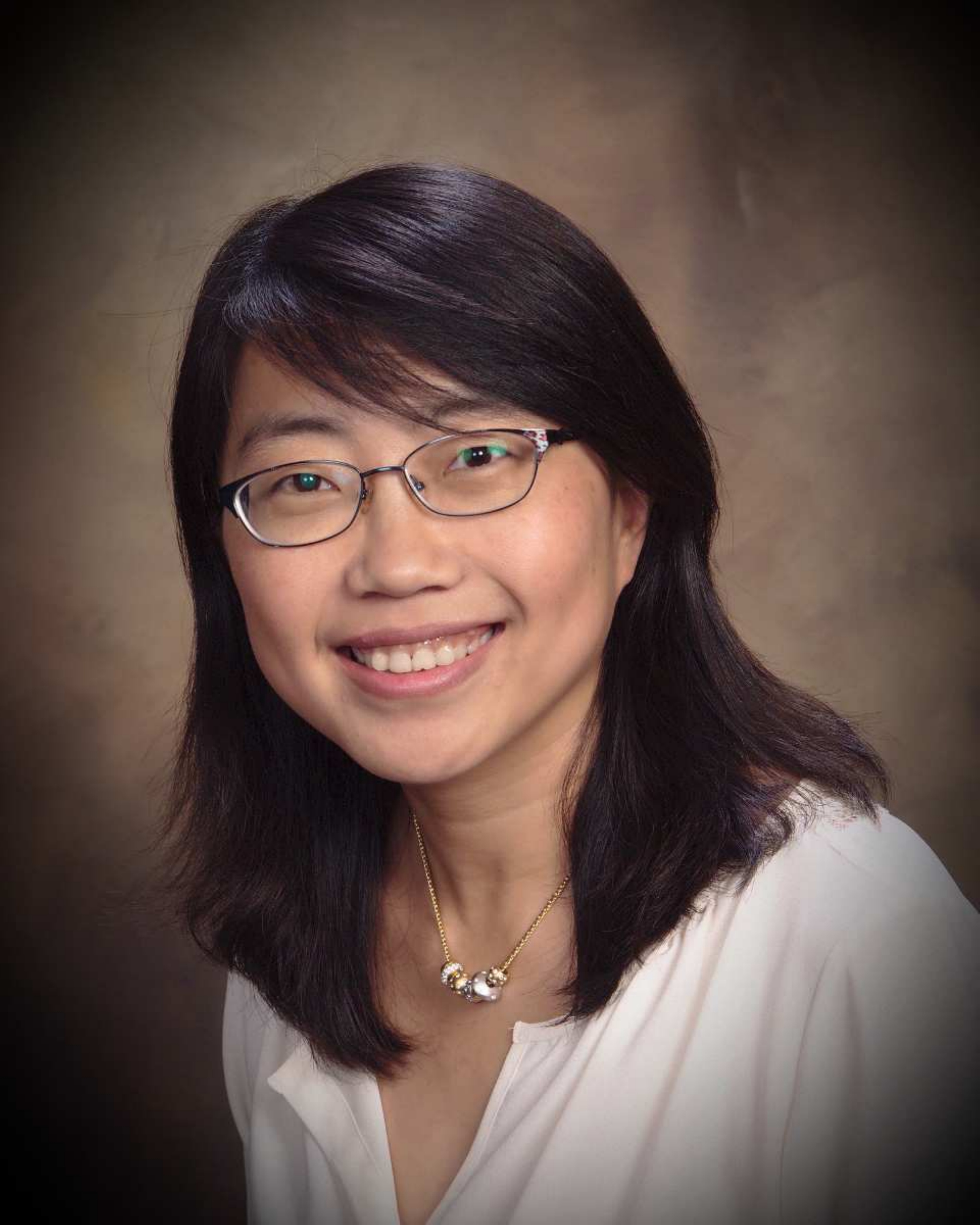}}]{Hai (Helen) Li}

(M'08-SM'16) received the B.S. and M.S. degrees from Tsinghua University, Beijing, China, and the Ph.D. degree from the Department of Electrical and Computer Engineering, Purdue University, West Lafayette, IN, USA.

Dr. Li currently is the Clare Boothe Luce Associate Professor with the Department of Electrical and Computer Engineering at Duke University, Durham, NC, USA. Prior to it, she was with Qualcomm Inc., San Diego, CA, USA, Intel Corporation, Santa Clara, CA, Seagate Technology, Bloomington, MN, USA, the Polytechnic Institute of New York University, Brooklyn, NY, USA, and the University of Pittsburgh, Pittsburgh, PA, USA. She has authored or co-authored more than 200 technical papers in peer-reviewed journals and conferences and a book entitled Nonvolatile Memory Design: Magnetic, Resistive, and Phase Changing (CRC Press, 2011). Her current research interests include memory design and architecture, neuromorphic architecture for brain-inspired computing systems, and architecture/circuit/device cross-layer optimization for low power and high performance.

Dr. Li serves as Associate Editor of IEEE TCAD, IEEE TVLSI, IEEE TCAS-II, IEEE TMSCS, ACM TECS, IEEE CEM, ACM TODAES, and IET-CPS. She was the General Chair or Technical Program Chair of multiple IEEE/ACM conferences and the Technical Program Committee members of over 30 international conference series. She received seven best paper awards and additional seven best paper nominations from international conferences. Dr. Li is a recipient of the NSF Career Award, DARPA Young Faculty Award (YFA), and TUM-IAS Hans Fisher Fellowship from Germany. She is a Distinguished Lecturer of the IEEE CAS society and a distinguished speaker of ACM. She is a senior member of the IEEE and a distinguished member of the ACM.

\end{IEEEbiography}

\vspace{-5mm}

\begin{IEEEbiography}[{\includegraphics[width=1in,height=1.25in,clip,keepaspectratio]{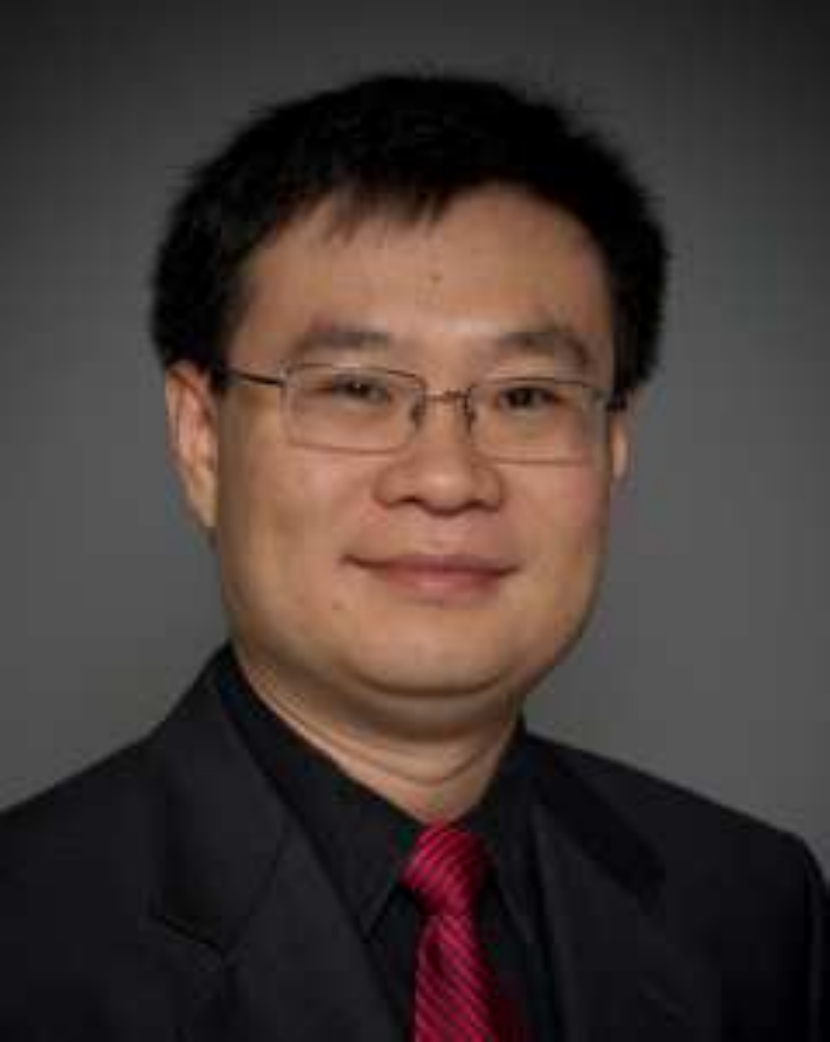}}]{Yiran Chen}

(M'04-SM'16-F'18) received B.S and M.S. from Tsinghua University and Ph.D. from Purdue University in 2005. After five years in industry, he joined University of Pittsburgh in 2010 as Assistant Professor and then promoted to Associate Professor with tenure in 2014, held Bicentennial Alumni Faculty Fellow. He now is a tenured Associate Professor of the Department of Electrical and Computer Engineering at Duke University and serving as the co-director of Duke Center for Evolutionary Intelligence (CEI), focusing on the research of new memory and storage systems, machine learning and neuromorphic computing, and mobile computing systems. Dr. Chen has published one book and more than 300 technical publications and has been granted 93 US patents. He is the associate editor of IEEE TNNLS, IEEE TCAD, IEEE D\&T, IEEE ESL, ACM JETC, ACM TCPS, and served on the technical and organization committees of more than 40 international conferences. He received 6 best paper awards and 14 best paper nominations from international conferences. He is the recipient of NSF CAREER award and ACM SIGDA outstanding new faculty award. He is the Fellow of IEEE.

\end{IEEEbiography}

\vspace{-5mm}

\begin{IEEEbiography}[{\includegraphics[width=1in,height=1.25in,clip,keepaspectratio]{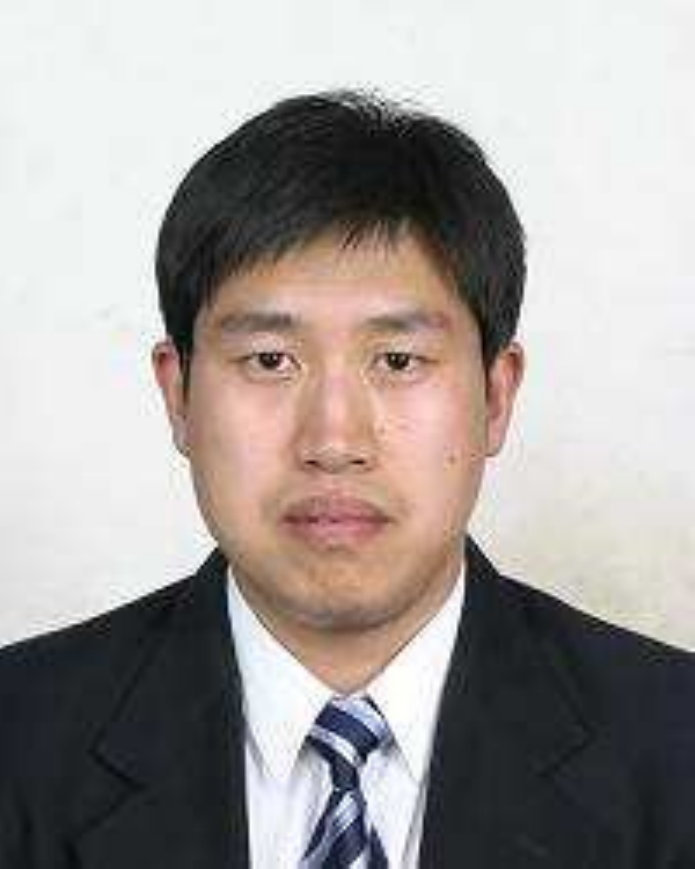}}]{Weisheng Zhao}

(M'06-SM'14) received the Ph.D. degree in physics from University of Paris Sud, Paris, France, in 2007.

He worked as a Research Associate at the CEA's embedded computing laboratory, France, from 2007 to 2009, and at the French national research center (CNRS), France, as a tenured scientist from 2009 to 2014 where he led the spintronics integration group. Now he is a professor and director of Fert Beijing Research Institute in Beihang University, Beijing, China. He has authored or coauthored 2 books, more than 200 scientific papers in the leading journals such as Nature Communications, Advanced Materials, Proceedings of the IEEE and he also holds 4 international patents and more than 50 Chinese patents.

Prof. Zhao is the associated editor of {\sc{IEEE Transactions on Nanotechnology}} and {\sc{IET Electronics Letters}}.

\end{IEEEbiography}

\end{document}